\documentclass[graybox]{svmult}

\usepackage{type1cm}        
\usepackage{makeidx}         
\usepackage{graphicx}        
\usepackage{mathrsfs}   
\usepackage{multicol}        
\usepackage[bottom]{footmisc}

\usepackage{color}

\usepackage{amsmath}
\usepackage{bm}
\usepackage{txfonts}
\renewcommand\mathbf{\bm}

\makeindex             

\begin{document}

\title*{QCD phase structure under rotation}
\author{Hao-Lei Chen, Xu-Guang Huang, and Jinfeng Liao}
\institute{Hao-Lei Chen \at Physics Department and Center for Particle Physics and Field Theory, Fudan University, Shanghai 200433, China,\\ \email{hlchen15@fudan.edu.cn}
\and Xu-Guang Huang \at Physics Department, Center for Particle Physics and Field Theory, and Key Laboratory of Nuclear Physics and Ion-beam Application (MOE), Fudan University, Shanghai 200433, China, \\ \email{huangxuguang@fudan.edu.cn}
\and Jinfeng Liao \at Physics Department and Center for Exploration of Energy and Matter, Indiana University, 2401 N Milo B. Sampson Lane, Bloomington, IN 47408, USA,\\ \email{liaoji@indiana.edu}}
%
%
\maketitle

\abstract{
We give an introduction to the phase structure of QCD matter under rotation based on effective four-fermion models. The effects of the magnetic field on the rotating QCD matter are also explored. Recent developments along these directions are overviewed, with special emphasis on the chiral phase transition. The rotational effects on pion condensation and color superconductivity are also discussed.}

\section {Introduction}\label{sec:intro}
\begin{figure}[!t]
\begin{center}
\includegraphics[scale=.2]{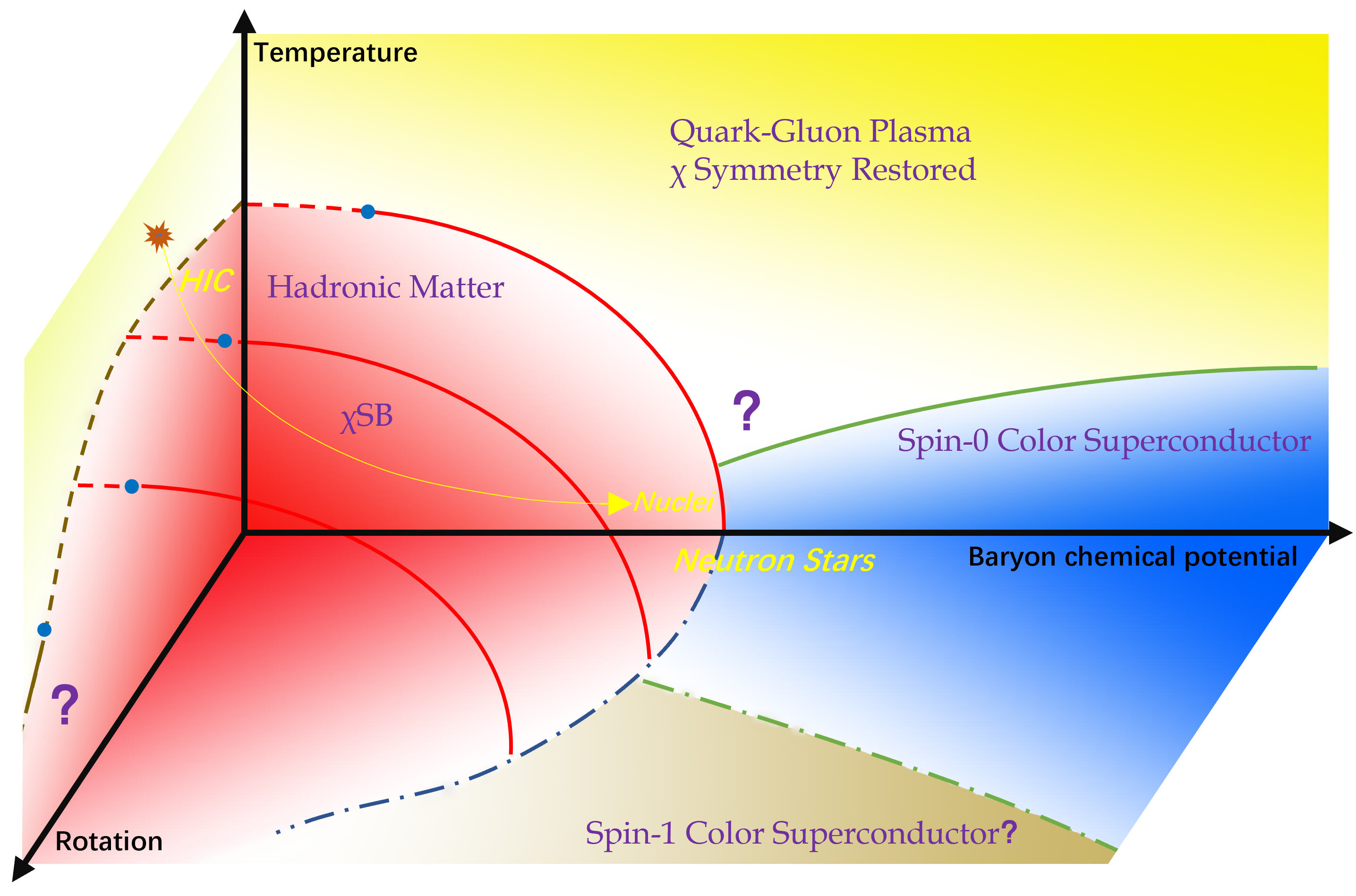}
\caption{A schematic phase diagram of QCD matter in the 3-dimensional parameter space spanned by the temperature$-$baryon chemical potential$-$rotation axes.}
\label{fig:illu}
\end{center}
\end{figure}

Exploration of the phase structure of quantum chromodynamics (QCD) is one of the most active researching frontiers of nuclear physics. In the past, most of the attention has been paid to the phase diagram in the plane of temperature $T$ and baryon density or baryon chemical potential $\mu_B$. At low temperature and low baryon chemical potential, the QCD matter is in confined hadronic phase where the (approximate) chiral symmetry of QCD Lagrangian is spontaneously broken. With increasing temperature QCD undergoes a transition from the hadronic matter to a deconfined and chirally symmetric state of quarks and gluons usually called the quark-gluon plasma (QGP). At zero $\mu_B$ this transition is not a true phase transition (i.e. without thermodynamic singularity) but a rapid crossover with the crossover region determined roughly by the confinement scale $\Lambda_{\rm  QCD}\sim 200$ MeV. However, at finite $\mu_B$, many model studies and theoretical arguments suggest that the restoration of  chiral symmetry should occur via a  first-order phase transition. The experimental search of the first-order phase transition at finite $\mu_B$ and its end point (i.e. the QCD critical end point) is one of the main goals of the RHIC beam energy scan program. At low $T$ but very high $\mu_B$, QCD is possibly in a color superconducting phase. This is confirmed at asymptotically high $\mu_B$ where the perturbative calculation shows that the ground state of QCD is a color-flavor-locking superconducting phase. At moderate $\mu_B$, we lack a first-principle calculation and the model studies suggest a series of possible color superconducting phases as well as other exotic phases. See, e.g. Ref.~\cite{Fukushima:2010bq} for review on QCD phase structure in $T-\mu_B$ plane, Refs.~\cite{Alford:2007xm} on color superconductivity, and Ref.~\cite{Bzdak:2019pkr} on the QCD critical end point and its experimental search.
 A schematic phase structure of QCD matter is shown in Fig.~\ref{fig:illu} where, in addition to the usual temperature and baryon chemical potential axes, a new dimension of finite global rotation is introduced. It is the influence of rotation on the QCD phase structures that is an emerging new direction of study and the main topic of our discussion.

The effects of the magnetic field on QCD phase structure have attracted a lot of interests in the past few decades. On one hand, this is because that the interplay between quantum electrodynamics (QED) and QCD proposes novel and interesting theoretical problems. On the other hand, this is also because that strong magnetic fields do exist in a wide range of physical systems usually governed by QCD physics, e.g. the neutron stars and heavy ion collision experiments. Recent theoretical studies have shown that the heavy-ion collisions generate the strongest magnetic fields (of order of $m_\pi^2$) in the current universe; see reviews~\cite{Hattori:2016emy,Huang:2015oca,Kharzeev:2015znc} and references therein. One remarkable consequence of the magnetic field is the magnetic catalysis of chiral condensate at zero temperature~\cite{Gusynin:1994re,Gusynin:1995nb}. Namely, the chiral condensate $\langle\bar{\psi}\psi\rangle(B)$ is generally enhanced compared with $\langle\bar{\psi}\psi\rangle(0)$ at $T=0$ where $B$ is an external magnetic field. This is understood as a result of dimensional reduction of charged-fermion dynamics in strong magnetic field. Surprisingly, the lattice QCD simulations revealed that the critical temperature for chiral phase transition of QCD is not enhanced but suppressed by the magnetic field~\cite{Bali:2011qj,Bali:2012zg}. This abnormal phenomenon is dubbed the inverse magnetic catalysis and is not fully understood yet. For reviews about the magnetic effects on QCD phase structure, see Refs.~\cite{Andersen:2014xxa,Miransky:2015ava}.

Recently, various properties of rotating QCD matter have become a new topic under active investigations. This was first inspired by the analogy between rotation and magnetic field. There is also a strong motivation from the discovery of extremely strong fluid vorticity structures (i.e. the local angular velocity or rotating frequency) in heavy-ion collisions where both  experiments~\cite{STAR:2017ckg,Adam:2018ivw,Adam:2019srw} and theories (see  e.g. Refs.~\cite{Deng:2016gyh,Jiang:2016woz,Shi:2017wpk,Deng:2020ygd}) find that the typical strength of the vorticity is about $10^{21}-10^{22}$ s$^{-1}$ (or tens of MeV) which may strongly influence the QCD matter. Furthermore, the rapidly rotating  pulsars (neutron stars) provide   another example of rotating QCD system, though the angular velocity is much smaller than the vorticity found in heavy-ion collisions. The rotation or fluid vorticity can polarize spin and thus induce a number of spin-related quantum phenomena. For example, it can induce a parity violating current called the chiral vortical effect (CVE)~\cite{Vilenkin:1979ui,Erdmenger:2008rm,Banerjee:2008th,Son:2009tf}, which is analogous to an effect (chiral magnetic effect) induced by magnetic field~\cite{Kharzeev:2007jp,Fukushima:2008xe}. It can also lead to detectable global polarization of  hadrons with nonzero spin (e.g.  $\Lambda$ hyperons and vector mesons) in heavy-ion collisions~\cite{Liang:2004ph,Liang:2004xn,Huang:2020rev,Gao:2020lxh,Becattini:2020sww,Liu:2020ymh,Becattini:2020ngo}.

It is  quite natural to ask what are the influence   of rotation on the QCD phase structure. This is, however, a hard question to answer, as the QCD phase transitions involve non-perturbative dynamics in general and the rotation further complicates the problem.  (A lattice simulation of QCD in rotating frame was given in Ref.~\cite{Yamamoto:2013zwa}, though the phase structure was not discussed.) In recent years, this question has been extensively studied by using effective models with   quarks or hadrons as dynamic degrees of freedom~\cite{Chen:2015hfc,Jiang:2016wvv,Ebihara:2016fwa,Huang:2017pqe,Chernodub:2016kxh,Chernodub:2017ref,Chernodub:2017mvp,Liu:2017zhl,Liu:2017spl,Zhang:2018ome,Wang:2018sur,Wang:2018zrn,Wang:2019nhd,Chen:2019tcp,Cao:2019ctl,Zhang:2020jux,Zhang:2020hct,Cao:2020pmm}.
It is found that in a uniformly rotating system at finite temperature, density, and magnetic field, angular velocity plays a role of an effective chemical potential for the angular momentum and its presence suppresses the spin-0 pairing of quarks~\cite{Chen:2015hfc,Jiang:2016wvv,Wang:2018sur}.
It is also confirmed that such a uniformly rotational effect on thermodynamics is invisible at zero temperature and density~\cite{Ebihara:2016fwa,Chernodub:2016kxh,Chernodub:2017ref}.
For nonuniform rotation, even the ground state can be affected by  the presence of rotation  and exhibit a vortex structure under sufficiently rapid rotation~\cite{Wang:2018zrn}.

Another very interesting question is the QCD phase structure when there are both rotation and magnetic field. Note that in both the heavy-ion collisions and the neutron stars, the strong rotation is accompanied with strong magnetic fields. Experimental measurements, showing a small difference between the hyperon and anti-hyperon spin polarization, are also indicative of the magnetic field and vorticity in a parallel configuration~\cite{Muller:2018ibh,Guo:2019joy,Guo:2019mgh}.  With a concurrent magnetic field, the  rotation is found to create an even richer  phase structure. For example, by using the Nambu--Jona-Lasinio (NJL) model, it was shown that under strong rotation the chiral condensate decreases with increasing  magnetic field   and eventually the chiral symmetry is restored. This phenomenon is named the ``rotational magnetic inhibition" as it is an analogy to the magnetic inhibition phenomenon in the finite density system~\cite{Chen:2015hfc}. These studies suggest a phase diagram of QCD in temperature-baryon chemical potential-rotation space as illustrated in Fig. \ref{fig:illu}, which we shall explain in the subsequent sections.

In the following, we start with an introduction on the rotating frame and the thermodynamic potential of NJL model in the rotating frame with or without the presence of a parallel magnetic field. We then discuss  the effects of the rotation with or without a parallel magnetic field on chiral condensate as well as on other types of scalar condensates. We use the natural unit with $\hbar=c=k_B=1$.

\section{Rotating Frame}\label{sec:frame}
Let us consider a cylindrical system which is rigidly rotating with angular velocity $\Omega$ ($\Omega>0$) in the inertial laboratory frame $\Sigma^\prime$ with the vector basis $(\hat\partial_{t'},\hat\partial_{x'},\hat\partial_{y'},\hat\partial_{z'})$. The Minkowski metric is:
\begin{equation}
  \eta_{\mu\nu}=\eta^{\mu\nu}=\mathrm {diag}(1,-1,-1,-1).
\end{equation}
We can go to the non-inertial rotating frame $\Sigma$ through the coordinate transformation (we assume the rotating axis is along $z$ axis)
\begin{equation}\label{eq:CTinCa}
\left\{
\begin{aligned}
x'  &= x\cos\Omega t-y\sin\Omega t \\
y'  &=  x\sin\Omega t+y\cos\Omega t \\
z'&= z \\
t'& = t
\end{aligned}
\right.
.
\end{equation}

From the coordinate transformation, it is easy to get the line element in the rotating frame
\begin{equation}
  ds^2=\eta_{\mu \nu}dx^{\prime\mu} dx^{\prime\nu}=g_{\mu \nu}dx^\mu dx^\nu=(1-\Omega^2r^2)dt^2+2\Omega ydxdt-2\Omega xdydt-dx^2-dy^2-dz^2,
\end{equation}
where $r=\sqrt{x^2+y^2}$. Thus the metric of the rotating frame is
\begin{equation}
\label{metric}
  g_{\mu\nu}=\left(
     \begin{array}{cccc}
       1-\Omega^2 r^2 &\Omega y& -\Omega x& 0 \\
       \Omega y & -1 & 0 & 0 \\
       -\Omega x & 0 & -1 & 0 \\
       0 & 0 & 0 & -1 \\
     \end{array}
   \right).
\end{equation}
The Christoffel symbol is given by
\begin{equation}
  \Gamma^\beta_{\mu \nu}=\frac{1}{2}g^{\beta \sigma}(\partial_\mu g_{\sigma \nu}+\partial_\nu g_{\sigma \mu}-\partial_\sigma g_{\mu \nu}),
\end{equation}
and the nonzero components are
\begin{equation}
  \Gamma^x_{tt}=-x\Omega^2,\quad \Gamma^x_{ty}=\Gamma^x_{yt}=-\Omega,\quad \Gamma^y_{tt}=-y\Omega^2,\quad \Gamma^y_{tx}=\Gamma^y_{xt}=\Omega.
\end{equation}
The Riemann curvature is identically zero.

The Lagrangian density of a complex scalar field in curved spacetime~\footnote{To specify the terminology, any non-Minkowski metric with either zero or nonzero Riemann curvature is referred to as a ``curved spacetime".} is
\begin{equation}
  \mathcal{L}=\sqrt{-g}[|D_\mu\phi|^2-(m^2+\xi R)|\phi|^2]=-\phi^* [\frac{1}{\sqrt{-g}}D_\mu(\sqrt{-g}g^{\mu\nu}D_\nu\phi)+(m^2+\xi R)\phi],
\end{equation}
where $m$ is the mass of the field, $g$ is the determinant of $g_{\mu\nu}$, and $\xi$ represents the coupling of the field with the Ricci curvature $R$ which is $0$ in the rotating frame. $D_\mu=\partial_\mu+iqA_\mu$ is the covariant derivative with $q$ the charge and $A_\mu$ the $U(1)$ gauge field. Note that the vector field should transform as $A_\mu(x)dx^\mu=A^\prime_\mu(x')dx^{\prime\mu}$ under coordinate transformation.
The Klein-Gordon equation can be derived from the Lagrangian density by the Euler-Lagrangian equation (for $R=0$),
\begin{equation}\label{eq:KGeq}
  \frac{1}{\sqrt{-g}}D_\mu(\sqrt{-g}g^{\mu\nu}D_\nu\phi)+m^2\phi=0.
\end{equation}

In order to discuss the spinor field in curved spacetime, it is convenient to use the vierbein formalism \cite{parker2009quantum,birrell1984quantum}.
The Dirac equation of free fermion in curved spacetime is
\begin{equation}\label{eq:cDriac}
[i\gamma^\mu(\partial_\mu+iqA_\mu+\Gamma_\mu)-m]\psi(x)=0,
\end{equation}
where $\gamma^\mu=e^\mu_i \gamma^i$, $e^\mu_i$ is the vierbein, which satisfies $g_{\mu\nu}=e^i_\mu e^j_\nu \eta_{ij}$, and $\Gamma_\mu$ is the spin connection given by
\begin{equation}\label{eq:spinconnection}
  \begin{split}
    &\Gamma_\mu=-\frac{i}{4}\omega_{\mu ij}\sigma^{ij},\\
    &\omega_{\mu ij}=g_{\alpha \beta}e^\alpha_i(\partial_\mu e^\beta_j+\Gamma^\beta_{\mu \nu}e^\nu_j),\\
    &\sigma^{ij}=\frac{i}{2}[\gamma^i,\gamma^j].
  \end{split}
\end{equation}
The Greek and the Latin letters denote the indices in coordinate and tangent (local Miknowski) spaces, respectively.

In the following, we will adopt
\begin{equation}
  e^t_0 = e^x_1 = e^y_2 = e^z_3 = 1 \;,\quad
  e^x_0 = y\Omega \;, \quad e^y_0 = -x\Omega \;,
\end{equation}
and other components are zero. From Eq.~(\ref{eq:spinconnection}), it is straightforward to get the only nonzero spin connection
\begin{equation}
    \Gamma_t=-i\Omega\sigma^{12}.
\end{equation}

\section{Nambu--Jona-Lasinio Model}\label{sec:NJL}
The Nambu--Jona-Lasinio (NJL) model is a four-fermion model which is commonly used as a low energy effective model of QCD~\cite{Klevansky:1992qe,Hatsuda:1994pi,Buballa:2003qv}. In curved spacetime, the NJL Lagrangian is
\begin{equation}\label{eq:NJL}
  {\mathcal L}_{\rm NJL}=\bar\psi(i\gamma^\mu\nabla_\mu-m_0)\psi+\frac{G}{2}[(\bar\psi\psi)^2+(\bar\psi i\gamma^5{\mathbf\tau}\psi)^2],
\end{equation}
where $\psi$ is the fermion field (representing the quarks), $\nabla_\mu=\partial_\mu + i \hat Q A_\mu +\Gamma_\mu$ is the covariant derivative, $\hat Q$ is the charge matrix in the flavor space, $G$ is the coupling constant, and $\mathbf \tau$ is the generator of the flavor group. For one flavor NJL model, $\mathbf\tau$ is just $1$. For two flavor NJL model, $\mathbf\tau$ should be the Pauli matrices of isospin $SU(2)$ group. In case that the quark current mass $m_0=0$, the NJL model has also chiral symmetry so that its symmetry group is $SU_R(2)\otimes SU_L(2)\otimes U_B(1)$, the same as that for QCD with two flavor quarks. When $m_0\neq0$ but small, the NJL model has an approximate chiral symmetry. Besides, the gauge color $SU(N_c)$ symmetry can be trivially assigned to Lagrangian (\ref{eq:NJL}) as an additional global symmetry. The generating functional of NJL model (with vanishing sources) is
\begin{equation}
\begin{split}
Z&=\int{\cal D} [\bar\psi,\psi] \exp\left(i\int d^4 x \sqrt{-g}{\mathcal L}_{\rm NJL}\right)\\
&=\int{\cal D} [\bar\psi,\psi, \sigma, \pi]\exp\left\{i\int d^4 x \sqrt{-g} \left[\bar\psi(i\gamma^\mu \nabla_\mu-m-i\gamma^5\pi\cdot\tau)\psi-\frac{\sigma^2+{\mathbf\pi}^2}{2G}\right]  \right\},
\end{split}
\end{equation}
where $m=m_0+\sigma$. In the second line, we have performed the Hubbard-Stratonovich transformation by introducing an auxiliary scalar field $\sigma$ and $N_\pi$ pseudoscalar fields $\mathbf\pi$ ($N_\pi$ is the number of the generators $\mathbf{\tau}$ of the flavor group, e.g. $N_\pi=3$ for two-flavor case). In the rest of this chapter, unless otherwise stated, we will consider only the rotating frame so that the metric is given by Eq. (\ref{metric}) and $g=-1$. Under mean field approximation, the one-loop effective action is
\begin{equation}\label{eq:effact}
  \Gamma=\frac{1}{i}\ln Z=-\int d^4x \frac{\sigma^2+\mathbf\pi^2}{2G}+\frac{1}{i}\ln \det (i\gamma^\mu\nabla_\mu-m-i\gamma^5\bm{\pi}\cdot\bm{\tau}).
\end{equation}
If we further assume that $\sigma$ and $\mathbf \pi$ are constant, the second term can be evaluated as
\begin{equation}
  \begin{split}
  \label{eq:effact2}
  &\frac{1}{i}\ln \det (i\gamma^\mu\nabla_\mu-m-i\gamma^5\mathbf\pi\cdot\mathbf\tau)=\frac{1}{2i}{\rm Tr}\ln[-(i\partial_t)^2+\hat H^2]\\
  &=\int dt\int \frac{dp_0}{2\pi}\sum_{\{\xi\}} \frac{1}{2i}\ln[-p_0^2+\varepsilon_{\{\xi\}}^2]
  \end{split}
\end{equation}
where
\begin{equation}\label{eq:Hamiltonian}
  \hat H=-i \gamma^0 \gamma^x\nabla_x-i \gamma^0 \gamma^y\nabla_y-i \gamma^0 \gamma^z\nabla_z+m \gamma^0+i \gamma^0 \gamma^5\mathbf\pi\cdot\mathbf\tau+\hat QA_t-i \Gamma_t,
\end{equation}
which we assume to be time independent, i.e. $[\hat H,i\partial_t]=0$, $\varepsilon_{\{\xi\}}$ is the eigenvalue of $\hat H$ with a set of quantum number $\{\xi\}$, and $p_0$ is the eigenvalue of $i\partial_t$. Then by employing Matsubara formalism
\begin{equation}
  \begin{split}
    t&\to-i\tau,\quad
p_0\to i\omega_n=i2\pi \frac{1}{\beta}(n+\frac{1}{2}),\quad
    \int\frac{dp_0}{2\pi}\to \frac{i}{\beta}\sum_{n},
  \end{split}
\end{equation}
where $\beta=1/T$ with $T$ the temperature, we can obtain the thermodynamic potential
\begin{equation}
  V_{\rm eff}=-i\frac{1}{\beta V}\Gamma=\frac{\sigma^2+{\mathbf \pi}^2}{2G}-\sum_{\{\xi\}}\left[\frac{\varepsilon_{\{\xi\}}}{2}+\frac{1}{\beta}\ln(1+e^{-\beta \varepsilon_{\{\xi\}}})\right].
\end{equation}
By minimizing the thermodynamic potential with respect to $\sigma$ and $\mathbf \pi$, we can get the gap equations
\begin{equation}
\label{gapequation}
  \frac{\partial V_{\rm eff}}{\partial \sigma}=0,\quad \frac{\partial V_{\rm eff}}{\partial \mathbf{\pi}}=\bm0.
\end{equation}
The stable thermodynamic state is given by the solution of the gap equations associated with the global minimum of $V_{\rm eff}$.

The above procedure can be easily modified to allow the condensates $\sigma$ and $\bm\pi$ to be inhomogeneous in space. But in this case $\hat H$ is technically very hard to be diagonalized, so we have to adopt certain approximation or perturbative expansion during the calculation. One approximation scheme is the {\it local density approximation} which assumes that the derivative of the condensate is negligible compare to the condensate itself ($\partial\sigma\ll \sigma^2$ or more precisely $\partial^n\sigma\ll \sigma^{n+1}$ with $n>0$; similarly for $\bm\pi$) so that we can treat the condensates as constant in solving the eigenvalue problem. Then Eq.~(\ref{eq:effact2}) is changed to
\begin{equation}
  \frac{1}{2i}\int d^4x\int\frac{d p_0}{2\pi}\sum_{\{\xi\}}\ln[-(p_0)^2+\varepsilon_{\{\xi\}}^2(x)]\Psi_{\{\xi\}}^\dagger\Psi_{\{\xi\}},
 \end{equation}
where $\Psi_{\{\xi\}}$ is the eigenfunction of $\hat H$ under the local density approximation. The thermodynamic potential is accordingly changed to
\begin{equation}
\label{therml}
  V_{\rm eff}=\frac{1}{\beta V}\int d^4 x_E\bigg\{\frac{\sigma^2+{\mathbf \pi}^2}{2G}-\sum_{\{\xi\}}\bigg[\frac{\varepsilon_{\{\xi\}}}{2}+\frac{1}{\beta}\ln(1+e^{-\beta\varepsilon_{\{\xi\}}})\bigg]\Psi_{\{\xi\}}^\dagger\Psi_{\{\xi\}}\bigg\},
\end{equation}
where $x_E^\mu$ is the Euclidean coordinate with compact time direction that is used in the Matsubara formalism. The gap equations are still given by Eqs.~(\ref{gapequation}).

\section{Rotating Fermions Without Boundary}\label{sec:unbound}
In this section, we will ignore the background gauge field and focus on the effect of rotation only. A uniformly rotating system must be finite so that the causality condition $\Omega R\leqslant 1$ (with $R$ the transverse size of the system) is satisfied. This requires appropriate boundary condition when solving the eigenvalue problem for $\hat{H}$. However, if we focus on the region far away from the boundary, we can ignore the influence of the boundary and take the $R\rightarrow\infty$ limit. We will consider this approximation in this section and examine the influence of the boundary in next section. To simplify the discussion, we further assume that $\bm\pi=\bm 0$ and $\sigma$ depends only on the transverse radius $r$.

The eigenvalue $\varepsilon_{\{\xi\}}$ of $\hat H$ is obtained by solving the Dirac equation under the local density approximation, which is given by (here $\{\xi\}= \{l, p_z, p_t, s\}$ with $s=\pm$ and we abbreviate $\varepsilon_{l, p_z, p_t, s}$ as $\varepsilon_{l,s}$) \cite{Jiang:2016wvv}
\begin{equation}\label{eq:ep_jiang}
  \varepsilon_{l,\pm}=\pm\sqrt{p_z^2+p^2_t+\sigma^2}-\Omega \left(l+\frac{1}{2}\right),
\end{equation}
where $p_z$ is the $z$-momentum, $p_t$ is the transverse momentum magnitude, and $l=0,\pm 1,\dots$ is the quantum number of the orbital angular momentum. We do not show the lengthy expression of the eigenfunction associated with $\varepsilon_{l,s}$; it is given in, e.g. Refs. \cite{Jiang:2016wvv,Ambrus:2014uqa,Ambrus:2019cvr}. From the dispersion relation~(\ref{eq:ep_jiang}), one can observe that rotation behaves very similar to a chemical potential, which has been noticed for a long time (see e.g. Ref. \cite{birrell1984quantum}).

It is worthy to compare the effect of rotation with the magnetic field $B$ on the dispersion relation. The latter gives the Landau levels:   $\varepsilon_{n,\pm}=\pm\sqrt{p_z^2+\sigma^2+2nq B}$.
In a background magnetic field, the transverse motion is quantized while there is no transverse-motion quantization in the rotation without boundary. This is because the existence of the centrifugal force due to rotation prevents the formation of a quantum mechanical bound-state problem. In the background magnetic field, each Landau level is highly degenerate with degeneracy $N=\lfloor qBS/(2\pi)\rfloor$ with $S$ the transverse area. As there is no transverse-motion quantization in the rotating case, we do not have such degeneracy. In fact, as we will show later, the Landau level degeneracy $N$ counts the number of allowed angular-momentum modes accommodated at each Landau level, which is lifted when there is a rotation. Thus the rotation behaves quite differently from the magnetic field.
\begin{figure}[t]
\begin{minipage}[t]{0.47\linewidth}
\includegraphics[scale=.13]{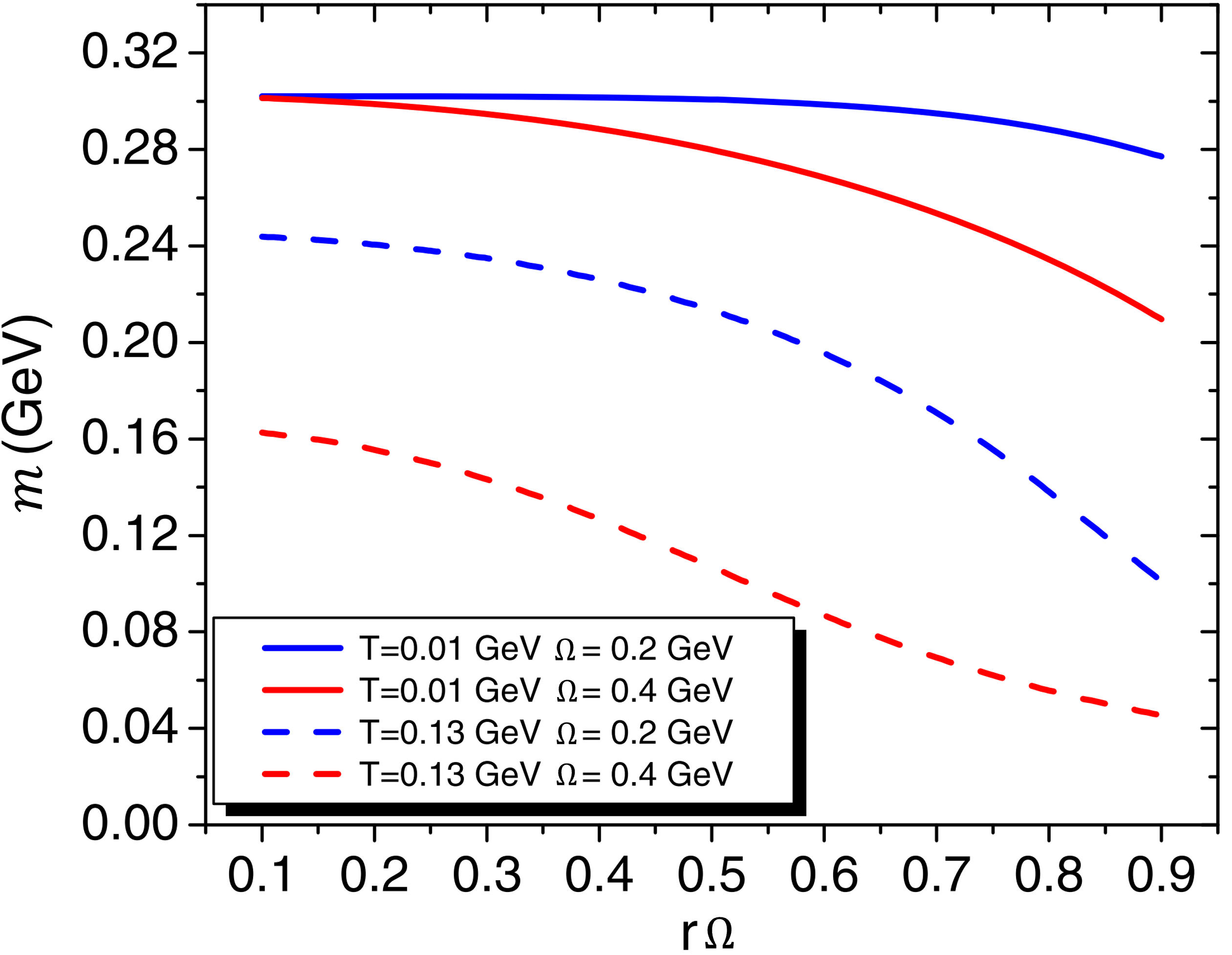}
\caption{The dependence of $m=\sigma+m_0$ on radial coordinate $r\Omega$ for several fixed values of $\Omega$ and $T$. (Taken from \cite{Jiang:2016wvv}.)}
\label{fig:jiang1}
\end{minipage}%
\hfill
\begin{minipage}[t]{0.47\linewidth}
\includegraphics[scale=.133]{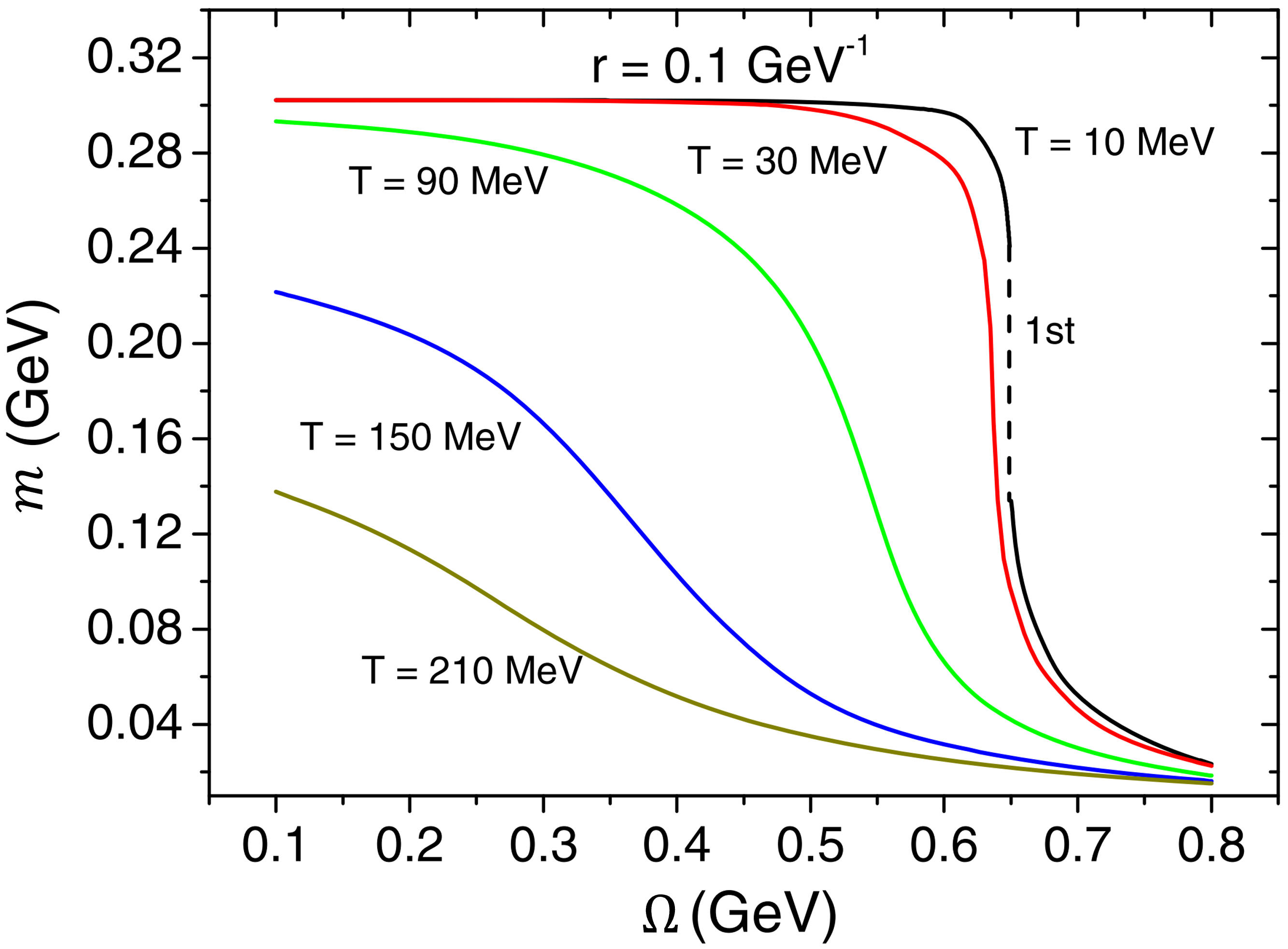}
\caption{Dependence of $m$ at radius $r=0.1$ GeV$^{-1}$ on $\Omega$ for various fixed value of $T$. (Taken from \cite{Jiang:2016wvv}.)}
\label{fig:jiang2}
\end{minipage}
\end{figure}

\begin{figure}[t]
\begin{minipage}[t]{0.47\linewidth}
\includegraphics[scale=.13]{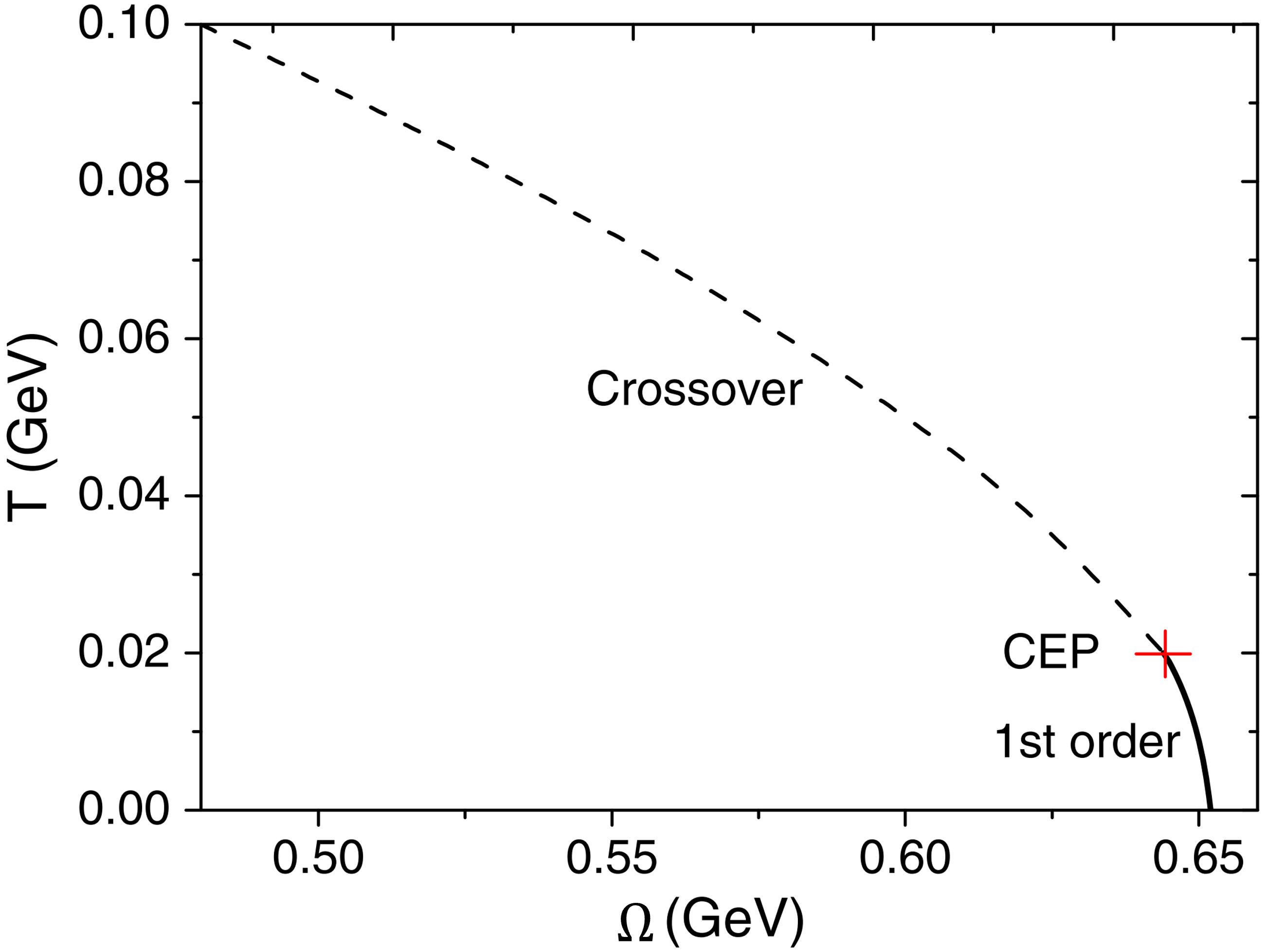}
\caption{The phase diagram on the $T-\Omega$ plane. (Taken from \cite{Jiang:2016wvv}.)}
\label{fig:jiang3}
\end{minipage}%
\hfill
\begin{minipage}[t]{0.47\linewidth}
\includegraphics[scale=.13]{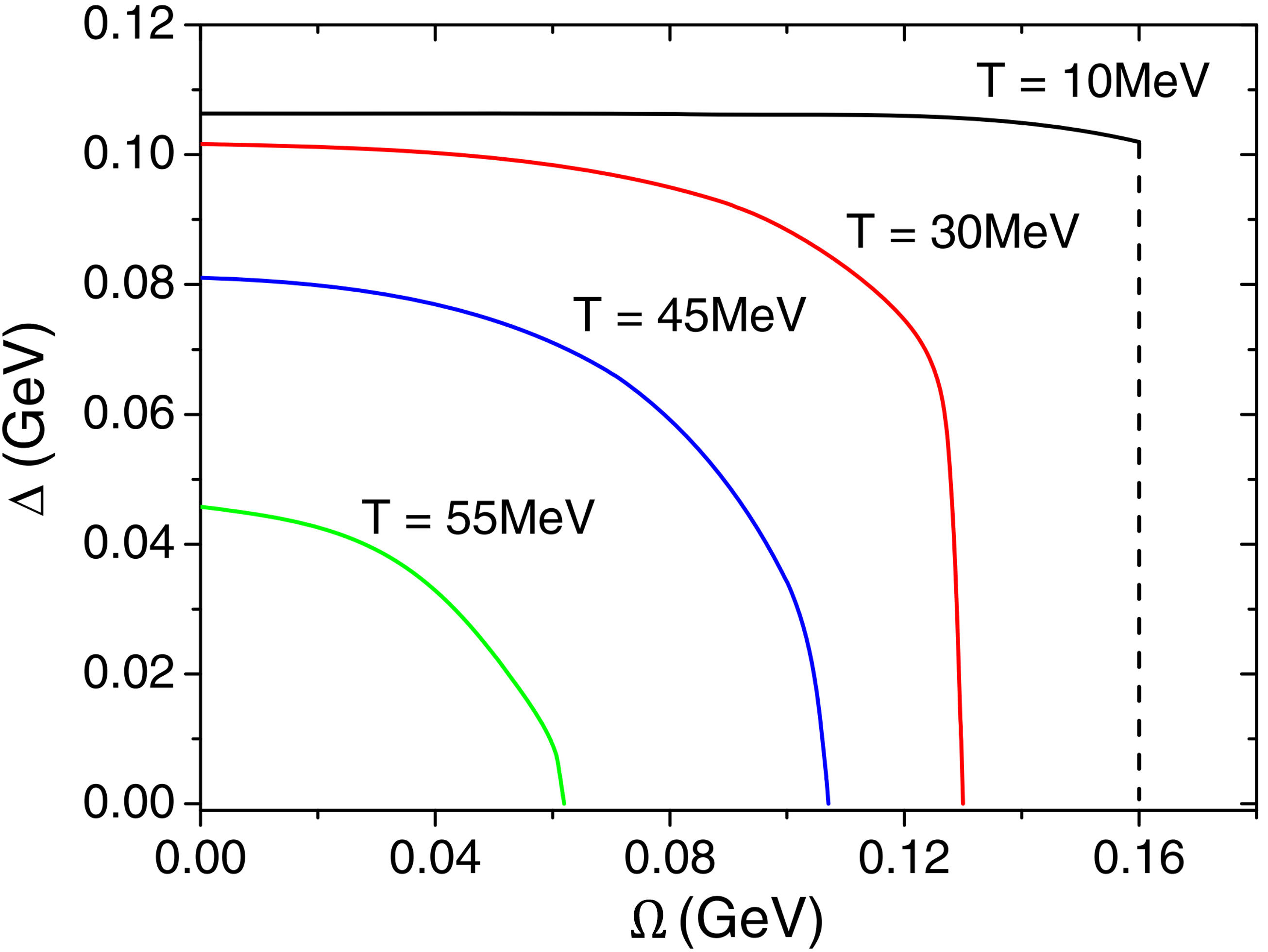}
\caption{The diquark condensate $\Delta$ at $r=0.1$ GeV$^{-1}$ as a function of $\Omega$ for several values of $T$ and fixed value of $\mu=\mu_B/3=400$ MeV. (Taken from \cite{Jiang:2016wvv}.)}
\label{fig:jiang4}
\end{minipage}
\end{figure}

Following the procedure introduced in the previous section, one can get the thermodynamic potential as
\begin{equation}
\begin{split}\label{eq:tp}
V_{\rm eff}=&\int d^3{\mathbf r}\Bigg\{\frac{\sigma^2}{2G}-\frac{2N_fN_c}{16\pi^2}\sum_l\int dp_t^2\int dp_z
  \\&\times T\ln(1+e^{\beta\varepsilon_{l}})(1+e^{-\beta\varepsilon_{l}})[J_l(p_tr)^2+J_{l+1}(p_tr)^2]
  \Bigg\},
\end{split}
\end{equation}
where $\varepsilon_{l}=\varepsilon_{l,+}$, $N_f=2$ is the flavor number, $N_c=3$ is the color number, and $J_l(x)$ is the Bessel function of the first kind. The condensate $\sigma$ is obtained from the gap equation. The numerical results are shown in Fig.~\ref{fig:jiang1} and Fig.~\ref{fig:jiang2} with model parameters given in \cite{Jiang:2016wvv}.
At all values of temperature, the mass gap $m$ (and thus the chiral condensate $\sigma$) decreases with increasing values of $\Omega$, which indicates the suppression effect of rotation on the chiral condensate. Furthermore, at low temperature the chiral condensate experiences a first-order transition when $\Omega$ exceeds a critical value $\Omega_c$, while at high temperature the chiral condensate vanishes with increasing $\Omega$ via a smooth crossover (it would be a second-order phase transition if $m_0=0$).

Figure \ref{fig:jiang3} is the $T-\Omega$ phase diagram obtained in \cite{Jiang:2016wvv}. We can see that the chiral symmetry is broken at low temperature and slow rotation. Qualitatively speaking, rotation will polarize the spin and orbital angular momentum of quarks along the direction of the angular velocity regardless of its charge, thus this polarization effect tends to destroy the pairing of the chiral condensate, which is total spin 0. If $\Omega$ is strong enough, the rotation would tend to forbid the formation of spin-0 pairing condensate and thus leads to a phase transition. For chiral condensate, this is very similar to the effect of baryon chemical potential $\mu_B$ which breaches the quark and anti-quark states and finally melts down the chiral condensate when $\mu_B$ is large enough. Hence, the chiral symmetry will be restored by increasing $T$ and/or $\Omega$. At high $T$ and low $\Omega$, there will be a smooth crossover. While at low $T$ and high $\Omega$, the transition is first-order. The crossover and the first-order transition is saperated by a critical end point. Fig.~\ref{fig:jiang3} is very similar to the $T-\mu_B$ phase diagram, which could be understood by considering $\Omega$ as a sort of ``chemical potential" for angular momentum.

Besides the chiral condensate, the authors of \cite{Jiang:2016wvv} also considered the diquark two-flavor superconducting (2SC) condensate at high density by adding the Lagrangian density (\ref{eq:NJL}) a chemical potential term and a diquark interaction
\begin{equation}
  {\mathcal L}_{d}=G_d(i\psi^TC \gamma^5 \psi)(i\psi^\dagger C \gamma^5 \psi^*),
\end{equation}
where $G_d$ is the diquark coupling constant and $C$ is the charge conjugation operator. Following similar procedure discussed in Sec. \ref{sec:NJL}, one can get the gap equation for the diquark condensate $\Delta\epsilon^{\alpha \beta 3}\epsilon_{ij}=-2G_d\langle i\psi^\alpha_iC \gamma^5\psi^\beta_j\rangle$ under the mean field approximation: $\partial V_{\rm eff}/\partial\Delta=0$. Their numerical result is shown in Fig~\ref{fig:jiang4}. We can see that the diquark condensate is also suppressed by the rotation simply because the 2SC pairing is also spin 0. Similarly, increasing $\Omega$ leads to a phase transition of melting of the diquark condensate, which is first order at low temperature while second order at high temperature.
\begin{figure}[t]
\begin{center}
\includegraphics[scale=0.22]{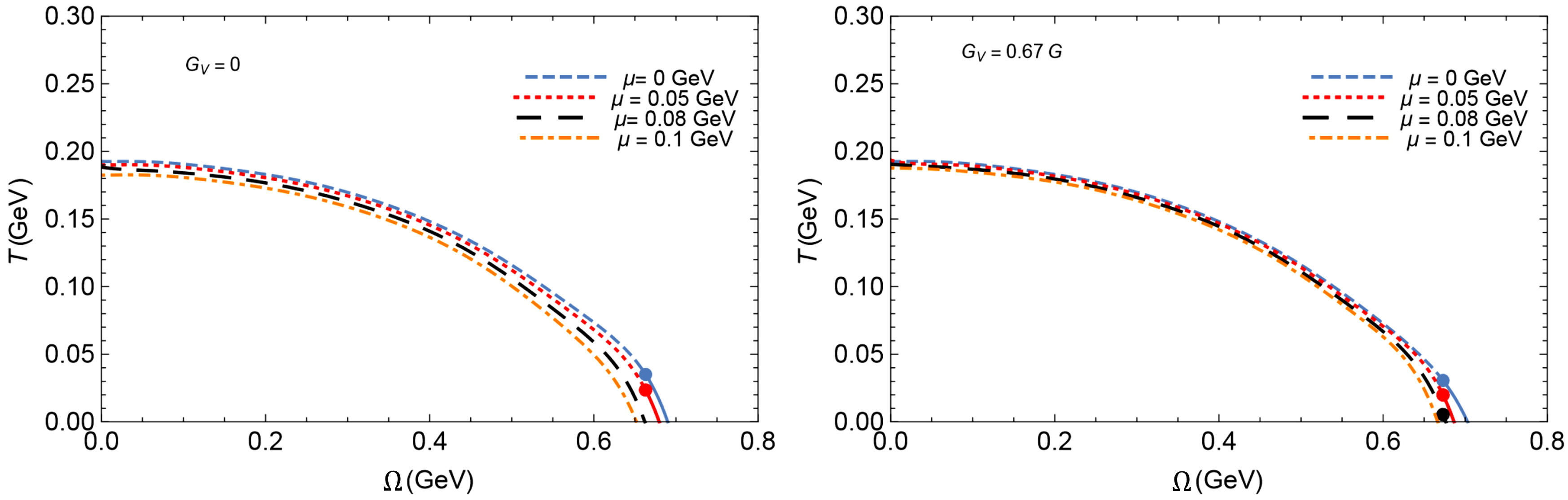}
\caption{The phase diagram in the $T-\Omega$ plane with different quark chemical potential $\mu$ for two different values of the vector coupling. (Taken from \cite{Wang:2018sur}.)}
\label{fig:mei1}
\end{center}
\end{figure}

Recently, the authors of \cite{Wang:2018sur} studied the influence of the rotation on chiral condensate with an additional vector channel interaction,
\begin{equation}\label{eq:Lv}
  \mathcal{L}_V=-(G_V/2)[(\bar\psi \gamma^\mu\psi)^2+(\bar\psi \gamma^\mu \gamma^5\psi)^2],
\end{equation}
where $G_V$ is the corresponding coupling constant.
In mean field approximation, the effective action becomes
\begin{equation}
  \Gamma=-\int d^4x \left[\frac{\sigma^2}{2G}-\frac{(\mu-\tilde\mu)^2}{2G_V}\right]+\frac{1}{i}\ln \det [(i\gamma^\mu\nabla_\mu-\sigma+\tilde\mu\gamma^0],
\end{equation}
where the effective quark chemical potential is defined as $\tilde\mu=\mu-G_V\langle \psi^\dagger\psi \rangle$ with $\mu$ the quark chemical potential. The thermodynamic potential is $V_{\rm eff}=-i\Gamma/(\beta V)$ and the two gap equations governing $\sigma$ and $\tilde{\mu}$ are $\partial V_{\rm eff}/\partial\sigma=\partial V_{\rm eff}/\partial\tilde{\mu}=0$.  Figure~\ref{fig:mei1} shows the $T-\Omega$ phase diagram with different chemical potentials. An interesting observation is that the increase of the chemical potential only shifts down the critical temperature $T_C$ and does not change the critical angular velocity much. Similar behavior happens in $T-\mu$ diagram with different $\Omega$, which again confirm the analogy between rotation and chemical potential.

\section{Boundary Conditions}\label{sec:bc}
As we stress in last section, due to the requirement of causality, the absolute value of velocity $v=\Omega r$ should not exceed the speed of light. Thus, for uniform rotation, the system size should be limited by
\begin{equation}
  \Omega R\leqslant 1.
\end{equation}
In the previous section, we assume that the system is large enough and the boundary effects on the bulk condensates can be ignored. In this section, we will discuss the influence of the boundary by considering two kinds of the boundary conditions, the no-flux boundary condition \cite{Ebihara:2016fwa,Ambrus:2015lfr} and the MIT bag boundary condition \cite{Ambrus:2015lfr,Chodos:1974je}; see Ref.~\cite{Ambrus:2019cvr} for more discussion on boundary conditions. Again, we consider a system with cylindrical symmetry so that the angular momentum is a good quantum number and we ignore the background gauge field. To make the discussions more transparent, we focus on $\sigma$ condensate only and take $m_0=0, N_f=1, N_c=1$ in this section.

The no-flux boundary condition requires no total incoming flux at the spatial boundary to keep the total charge constant in the cylinder,
\begin{equation}\label{eq:noflux}
  \int d\theta\bar\psi\gamma^r\psi\Big |_{r=R}=0,
\end{equation}
where $\gamma^r=\gamma^1\cos\theta+\gamma^2\sin\theta$. One important feature of the no-flux boundary condition is that it can guarantee the Dirac Hamiltonian to be Hermitian. Due to the boundary condition, the transverse momentum should take discrete values, so we will denote it as $p_{l,k}$. Since the condition Eq.~(\ref{eq:noflux}) does not uniquely fix the solution of the Dirac equation, further requirement should be imposed \cite{Ambrus:2015lfr}. For example,
\begin{equation}
\label{condition}
  p_{l,k}=
\left\{
\begin{aligned}
&\xi_{l,k}R^{-1} \quad \text{for} \quad l=0,1,\dots \\
&\xi_{-l-1,k}R^{-1} \quad \text{for} \quad l=-1,-2,\dots  \\
\end{aligned}
\right.
\end{equation}
where $\xi_{l,k}$ represents the $k$th zero of the Bessel function $J_l(x)$. By employing the local density approximation, the gap equation at $T=\mu=0$ reads \cite{Ebihara:2016fwa}
\begin{equation}\label{eq:gap_ebi}
\begin{split}
\sigma=\frac{\sigma}{4G}\int^{+\infty}_{-\infty} dp_z\sum_{l=-\infty}^{\infty}\sum_{k=1}^{\infty}
\frac{2}{[J_{l+1}(p_{l,k}R)]^2R^2}\frac{J_{l}(p_{l,k}r)^2+J_{l+1}(p_{l,k}r)^2}
{E}\theta(E-|\Omega j|),
\end{split}
\end{equation}
where $E=\sqrt{p_{l,k}^2+p_z^2+\sigma^2}$ and $j=l+1/2$. The expression of the gap equation is very similar to the finite density system, the effect of rotation appears only in the theta function with $|\Omega j|$ playing a role of an effective chemical potential here. Therefore, the rotational effect appears only when $E<|\Omega j|$ for some $j$. If we do not consider the boundary condition, the transverse momentum $p_{l,k}$ takes continuous value from $0$ to $+\infty$. One can find a region of transverse momentum to satisfy $E<|\Omega j|$. However, once we take the boundary condition ~(\ref{eq:noflux}) into account, there is no mode that satisfies $E<|\Omega j|$. In fact, $E$ is minimized by setting $p_z=\sigma=0$ and $k=1$. By using an inequality for the zeros of the Bessel function \cite{Giordano:1983bes,Ebihara:2016fwa}
\begin{equation}
\begin{split}
\xi_{l,1}&>l+1.855757l^{1/3}+0.5l^{-1/3}\quad \text{for} \quad l\geqslant 1,\\
\xi_{0,1}&=2.40493>1/2,
\end{split}
\end{equation}
and the causality constraint $\Omega R \leqslant1$, one has the inequility for $l\geqslant 0$
\begin{equation}
  E-\Omega j\geqslant\frac{1}{R}(\xi_{l,1}-\Omega Rj)\geqslant\frac{1}{R}(\xi_{l,1}-j)>0.
\end{equation}
In the same way, for $l<0$, one can also prove that $E>|\Omega j|$. Thus uniform rotation has no effect in vacuum with no-flux boundary condition. We can understand this fact by comparing to the finite density system. In finite density system, the effect of chemical potential will be visible only when it exceed the mass threshold, which is known as the Sliver Blaze problem in finite density system. In rotating system, the effective chemical potential $|\Omega j|$ can never exceed the threshold $p_{l,1}$, thus the uniform rotation cannot induce a visible effect in vacuum.

\begin{figure}[t]
\begin{center}
\includegraphics[scale=0.14]{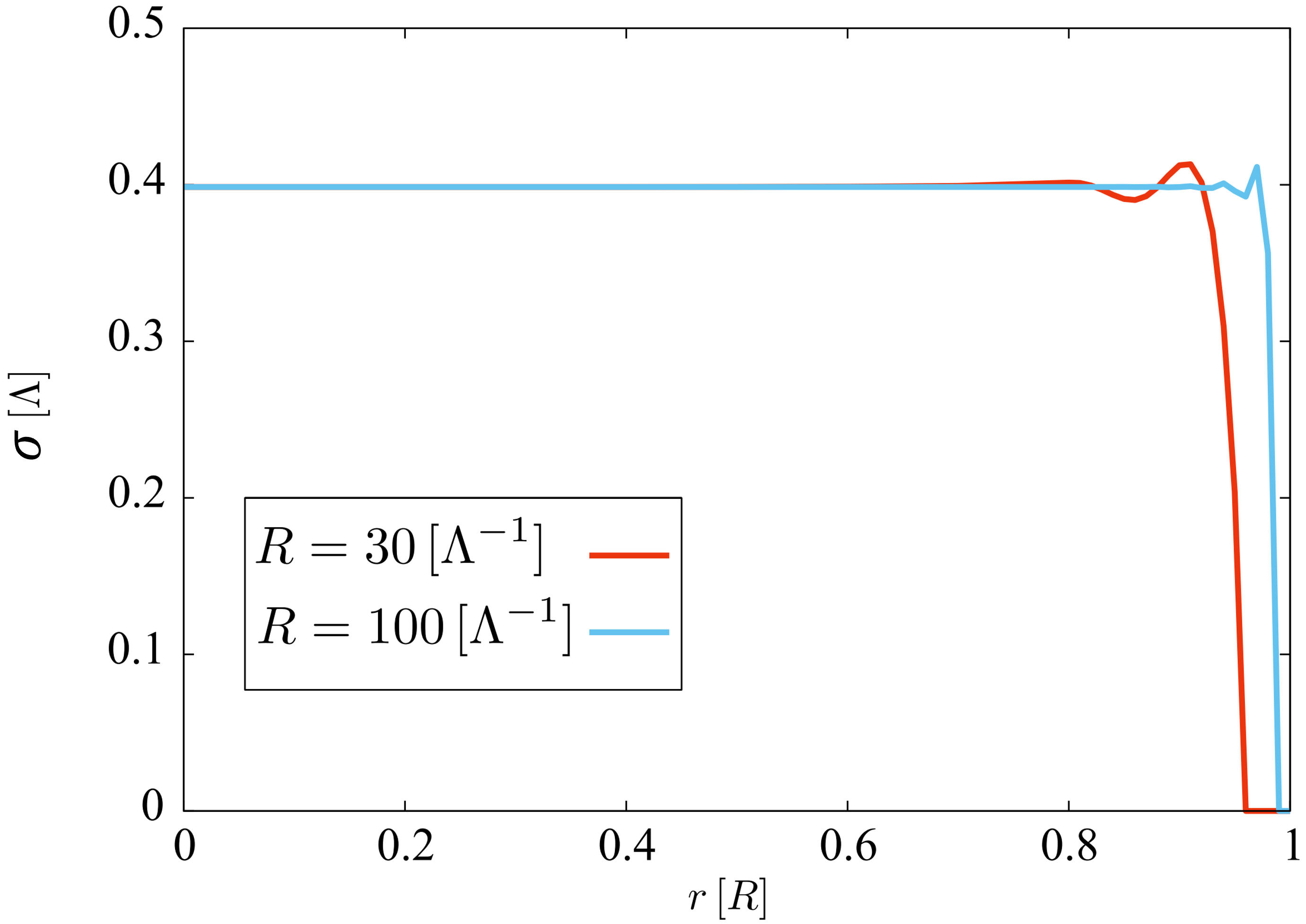}{}
\caption{Inhomogeneous chiral condensate $\sigma$ as a function of the radial coordinate $r$. (Taken from \cite{Ebihara:2016fwa}.)}
\label{fig:ebihara1}
\end{center}
\end{figure}

Although we do not have any rotational effect at $T=\mu=0$, it is still worthwhile to see the finite-size effect with the no-flux boundary condition. By numerically solving the gap equation (\ref{eq:gap_ebi}), one can get the result as shown in Fig.~\ref{fig:ebihara1}~\cite{Ebihara:2016fwa}. One can observe that the local density approximation is invalid in the vicinity of the boundary. The oscillation comes from the cutoff $\Lambda$: As the NJL model is not renormalizable, an ultraviolet cutoff must be introduced. As $R$ increases, the oscillation behavior becomes milder. The vanishing
condensate at the boundary is a consequence of the condition (\ref{condition}).

In \cite{Chernodub:2016kxh}, the authors adopted another boundary condition, the MIT boundary condition
\begin{equation}\label{eq:MIT}
  [i \gamma^\mu n_\mu(\theta)-1]\psi\Big|_{r=R}=0,
\end{equation}
where $n_\mu(\theta)=(0,\cos\theta,-\sin \theta,0)$ is a unit vector normal to the cylinder surface. It is easy to check that the MIT boundary condition Eq.~(\ref{eq:MIT}) leads to
\begin{equation}
   j^\mu n_\mu=0 \quad \text{at} \quad r=R,
\end{equation}
where $j^\mu=\bar\psi\gamma^\mu\psi$ is the current. This also leads to $\bar\psi \psi=0$ at the boundary. Thus we have the current vanishing at any point on the surface of the cylinder, a condition that is stronger than the no-flux boundary condition~(\ref{eq:noflux}).
\begin{figure}[!t]
\begin{center}
\includegraphics[scale=0.23]{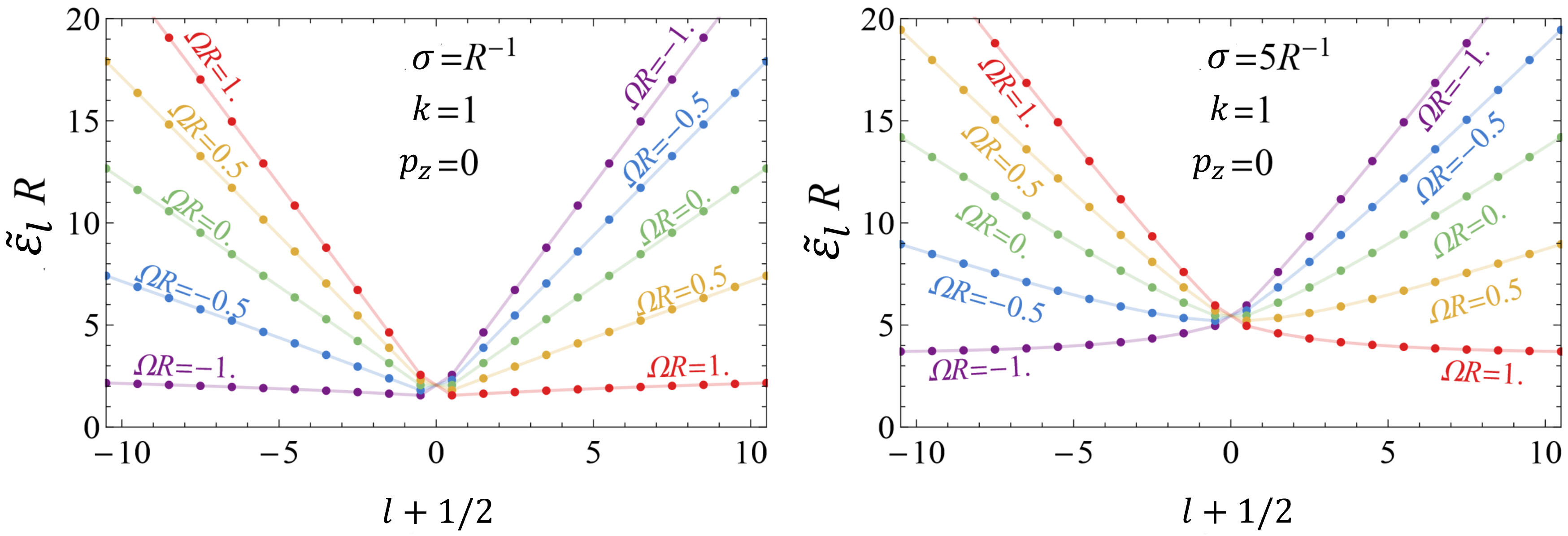}{}
\caption{Lowest energy eigenmodes with $k=1$ and $p_z=0$ versus
the angular momentum $j=l+1/2$ for various values of the rotation frequency $\Omega$. (Taken from \cite{Chernodub:2016kxh}.)}
\label{fig:chernodub1}
\end{center}
\end{figure}

By solving the Dirac equation (\ref{eq:cDriac}) with $m=\sigma$ with the MIT boundary condition, the discrete transverse momentum $p_{l,k}$ is given by the $k$th positive root of
\begin{equation}\label{eq:MITeq}
  j^2_l(p_{l,k}R)+\frac{2 \sigma }{p_{l,k}}j_l(p_{l,k}R)-1=0,
\end{equation}
where
\begin{equation}
  j_l(x)=\frac{J_l(x)}{J_{l+1}(x)}.
\end{equation}
The lowest energy spectrum $\tilde{\varepsilon}_l\equiv\varepsilon_{l,1,+}(p_z=0)$ with $\varepsilon_{l,k,\pm}=\pm\sqrt{p_{l,k}^2+p_z^2+\sigma^2}-\Omega(l+1/2)$ is shown in Fig.~\ref{fig:chernodub1} \cite{Chernodub:2016kxh}. It is easy to prove that the change of the orbital number $l\to -l-1$ (i.e. $j\to -j$) will not affect the eigenvalue $p_{l,k}$ in Eq. (\ref{eq:MITeq}):
\begin{equation}
  p_{l,k}=p_{-l-1,k},
\end{equation}
which is a result of $CP$ symmetry of the non-rotating system. Thus we can see that energy spectrum is doubly degenerated in Fig. \ref{fig:chernodub1} at $\Omega=0$. The spectrum will become asymmetric at $\Omega\neq 0$ because the rotation will explicitly break the $CP$ symmetry. However, the spectrum is invariant under the simultaneous flips $j\to -j$ and $\Omega \to -\Omega$.

An interesting observation of Fig. \ref{fig:chernodub1} is that $\tilde{\varepsilon}_l$ is always positive indicating the inequality
\begin{equation}
  E> |\Omega j|
\end{equation}
to hold also for MIT boundary condition. In fact, this inequality can be proven similarly as for the no-flux boundary condition. Thus one again confirms that the uniform rotation have no effect on the chiral condensate in the vacuum. In other words the cold vacuum does not rotate \cite{Chernodub:2016kxh}.

\begin{figure}[t]
\begin{minipage}[t]{0.47\linewidth}
\includegraphics[scale=.12]{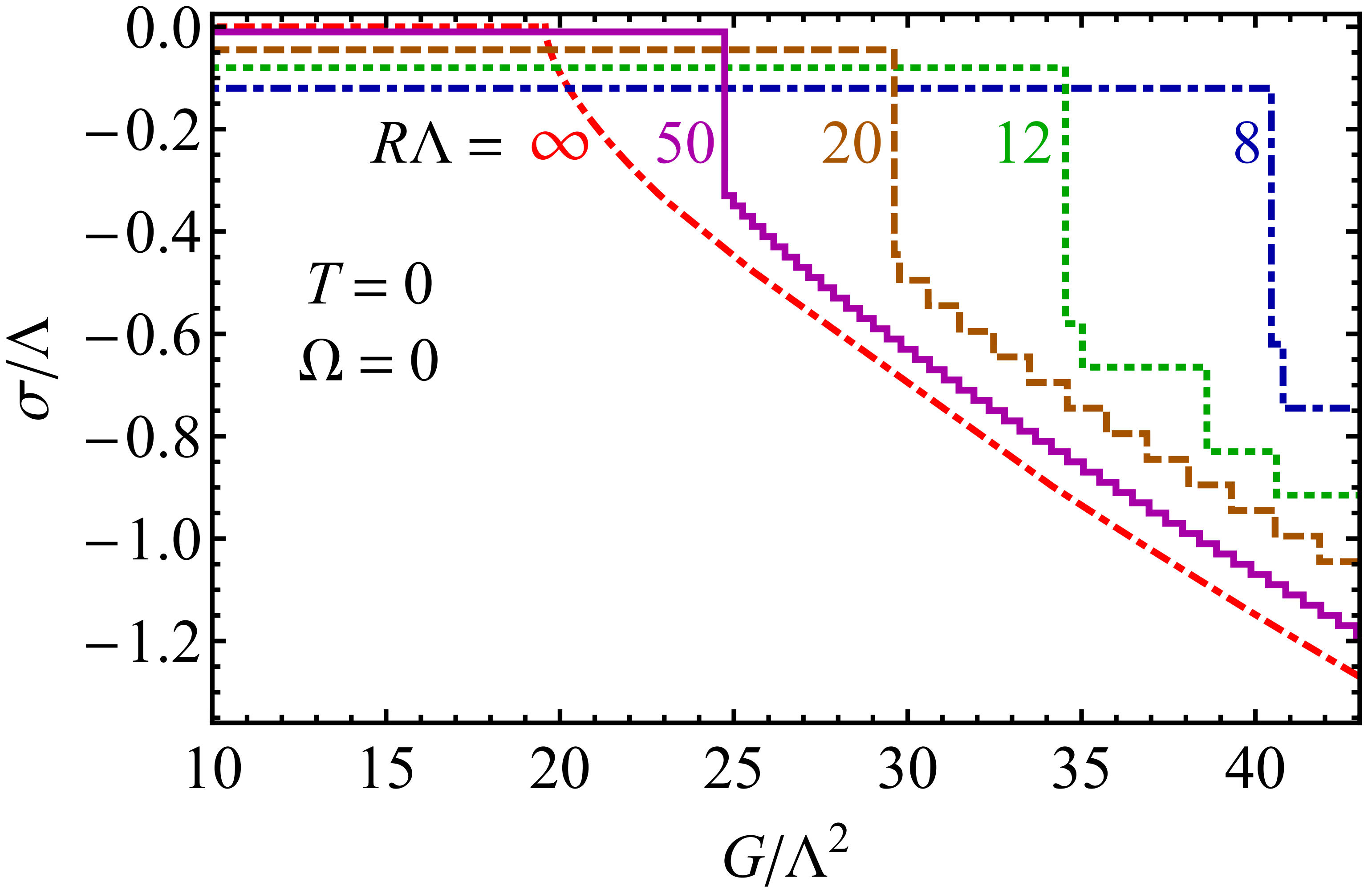}
\caption{The ground-state condensate
$\sigma$ in a non-rotating cylinder as a function of the coupling constant $G$ at various fixed radii $R$. Here, $\Lambda$ is the ultraviolet cutoff for NJL model. (Taken from \cite{Chernodub:2016kxh}.)}
\label{fig:chernodub2}
\end{minipage}%
\hfill
\begin{minipage}[t]{0.47\linewidth}
\includegraphics[scale=.124]{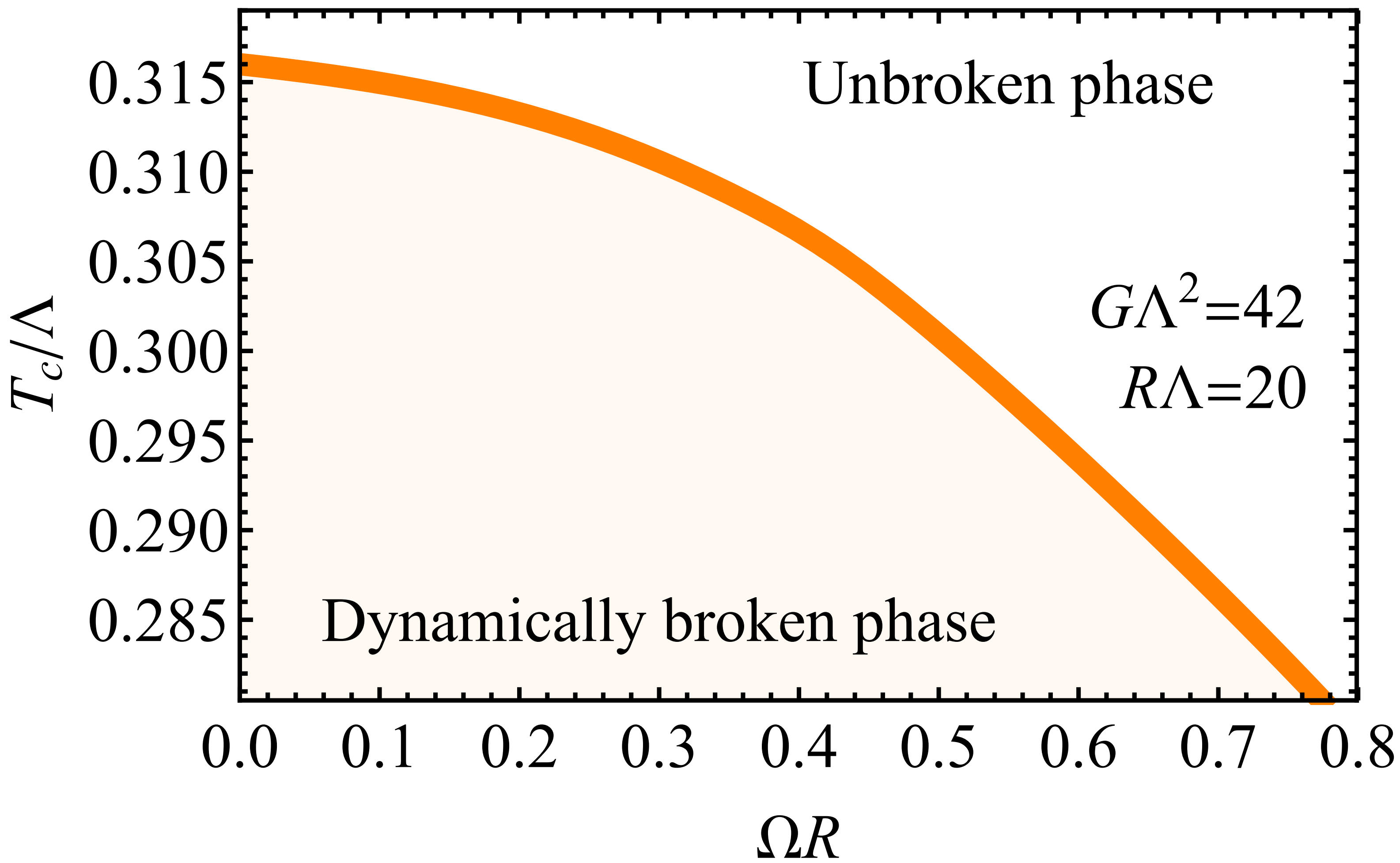}
\caption{The phase diagram of the rotating fermionic matter in the $T-\Omega$ plane with MIT boundary condition. (Taken from \cite{Chernodub:2016kxh}.)}
\label{fig:chernodub3}
\end{minipage}
\end{figure}

Then let's have a look at how the MIT boundary condition affects the chiral condensate at zero temperature and chemical potential. In \cite{Chernodub:2016kxh}, the authors assumed $\sigma$ to be homogeneous and computed $\sigma$ as a function of the coupling constant, as shown in Fig. \ref{fig:chernodub2}. We can see that the negative condensate is favored which has multiple steplike discontinuities, similar to the Shubnikov-de Haas oscillation, due to the discretization of the excitation levels. If we take $R\to \infty$, the discontinuities will disappear. It should be mentioned that the MIT boundary condition explicitly breaks the chiral symmetry. Thus at small $R$, the boundary effect is strong and we have large $\sigma$. While at large $R$, the boundary effect becomes weak, and the explicit breaking of chiral symmetry can be ignored.

At finite temperature, the rotational effect becomes visible. The phase diagram in $T-\Omega$ plane is obtained in \cite{Chernodub:2016kxh} and shown in Fig. \ref{fig:chernodub3}. The rotation in a finite cylinder tends to restore the chiral symmetry which is in agreement with the result in the previous section. However, the chiral phase transition is first order and is steplike due to the boundary condition. We note that in the restored phase, the condensate is still nonzero but has a small value due to the explicit breaking of the chiral symmetry by the MIT boundary condition.

\begin{figure}[t]
\begin{minipage}[t]{0.47\linewidth}
\includegraphics[scale=.13]{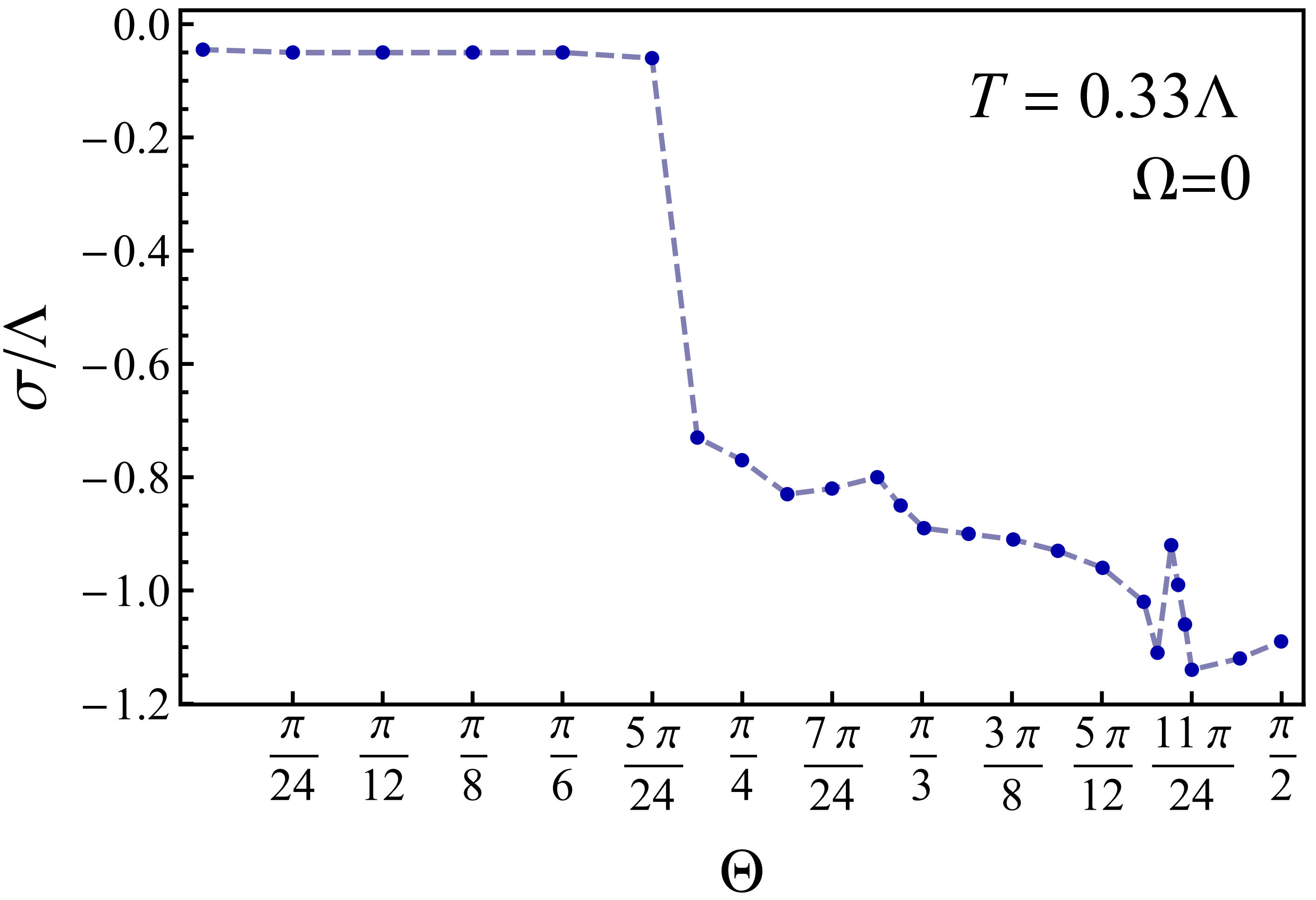}
\caption{The condensate $\sigma$ in vacuum in the non-rotating
cylinder as a function of the chiral angle $\Theta$. The cylinder radius is $R=20/\Lambda$ and the coupling $G =42\Lambda^2$. (Taken from \cite{Chernodub:2017ref}.)}
\label{fig:gongyo1}
\end{minipage}%
\hfill
\begin{minipage}[t]{0.47\linewidth}
\vspace{-3.25cm}
\includegraphics[scale=.135]{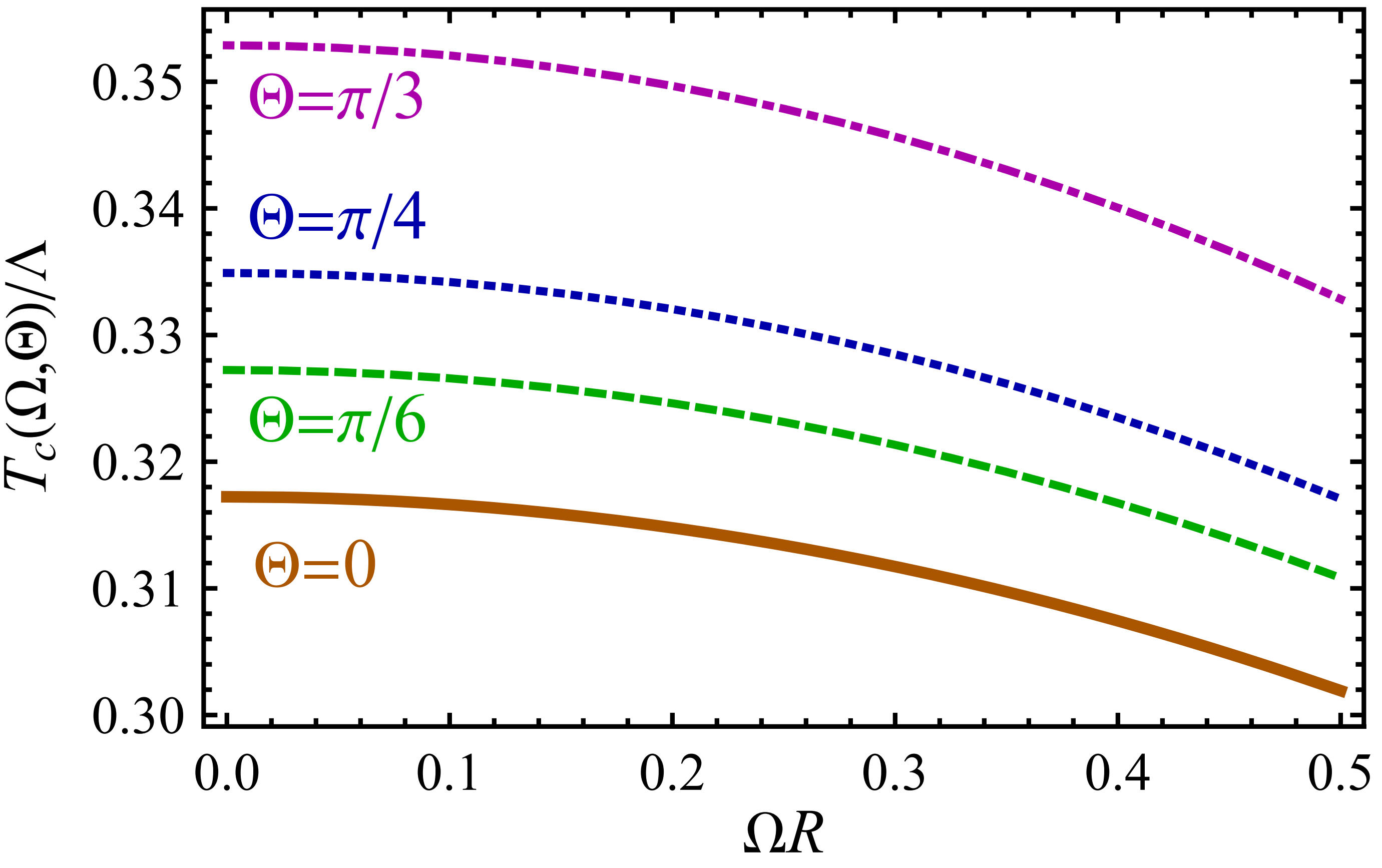}
\caption{Phase diagram of the rotating fermionic matter in
$(T,\Omega)$ plane at different angles $\Theta$. The symmetry breaking phase
is at lower part of the diagram. (Taken from \cite{Chernodub:2017ref}.)}
\label{fig:gongyo2}
\end{minipage}
\end{figure}

The MIT boundary condition can be generalized to \cite{Chernodub:2017ref}
\begin{equation}\label{eq:cMIT}
  [i \gamma^\mu n_\mu(\theta)-e^{-i\Theta\gamma^5}]\psi\Big|_{r=R}=0,
\end{equation}
where $\Theta$ is a chiral angle which parametrizes the chiral boundary condition. Then Eq.~(\ref{eq:MITeq}) becomes
\begin{equation}\label{eq:cMITeq}
j^2_l(p_{l,k}R)+\frac{2 \sigma }{p_{l,k}}j_l(p_{l,k}R)\cos\Theta-1=0.
\end{equation}
The usual MIT boundary condition corresponds to $\Theta=0$. One special choice of the chiral angle is $\Theta=\pi$, which only filp the sign of the mass term, $\sigma\to -\sigma$ in Eq.~(\ref{eq:MITeq}). Thus one can get all the previous result of MIT boundary condition only with the sign filp in the condensate $\sigma\to -\sigma$. Another special choice is $\Theta=\pi/2$. In this case, the values of $p_{l,k}$ are independent of the mass. Since the spectrum is affected by the choice of the chiral angle $\Theta$, one can expect that the phase diagram will exhibit a certain dependence on $\Theta$. It is easy to prove that the thermodynamic potential has following properties
\begin{equation}\label{eq:veffmit}
\begin{split}
 V_{\rm eff}(\sigma,\Theta)=V_{\rm eff}(-\sigma,\pi-\Theta)=V_{\rm eff}(\sigma,2\pi-\Theta).
\end{split}
\end{equation}
Thus one only need to consider the interval $\Theta\in[0,\pi/2]$ while other values of $\Theta$ can be restored from Eq.~(\ref{eq:veffmit}). In Fig.~\ref{fig:gongyo1} \cite{Chernodub:2017ref}, we can see how the boundary condition (\ref{eq:cMIT}) affects the chiral condensate in a non-rotating cylinder at finite temperature. At low values of the boundary angle $\Theta$, the system resides in a phase with a dynamically unbroken chiral symmetry in which a weak, explicit, violation of the chiral symmetry occurs. The explicit breaking is caused by the fact that the boundary conditions are not invariant under the chiral transformation as mentioned above. At $\Theta=\Theta_c\approx 5\pi/24$, the chiral condensate suddenly changes from a small negative value to a larger negative value, i.e. a first order transition occurs.
The phase diagram in the temperature-rotation $(T,\Omega)$ plane for various angles of the boundary chiral angle $\Theta$ is shown in Fig.~\ref{fig:gongyo2}. From the similarity of all these phase lines, one can infer that the effect of rotation is always tend to restore the chiral symmetry. On the other hand, the critical temperature of the chiral phase transition depends substantially on the boundary condition.

We note that besides the real solutions, Eq.~(\ref{eq:MITeq}) also has purely imaginary solutions~\cite{Chernodub:2017mvp}. These solutions are corresponding to the ``edge states" because their wave functions are localized at the boundary.

\section{Rotating Fermions with Background Magnetic Field}\label{sec:OmegaB}

In this section, we will take into account the background magnetic field. And we will see that there will be some interesting effects caused by the combination of rotation and magnetic field. We consider chiral limit $m_0=0$ in this section.

For simplicity, we introduce a constant magnetic field in the inertial lab frame $\Sigma^\prime$ along the $z'$-axis (which coincides with $z$-axis in the rotating frame) and assume $q{\mathbf B}\cdot{\mathbf\Omega}>0$. To preserve the rotational symmetry, we choose the symmetric gauge
\begin{equation}
    A^\prime_\mu=(0,\frac{1}{2}By',-\frac{1}{2}Bx',0).
  \end{equation}
Making a coordinate transformation to the rotating frame, we obtain the gauge vector in the rotating frame:
\begin{equation}
   A_\mu=(-\frac{1}{2}B\Omega r^2,\frac{1}{2}By,-\frac{1}{2}Bx,0).
 \end{equation}
From the Klein-Gordon equation (\ref{eq:KGeq}) and the Dirac equation (\ref{eq:cDriac}), the dispersion relation is given by~\cite{Chen:2015hfc,Mameda:2015ria}
\begin{equation}\label{eq:epB}
   [\varepsilon+\Omega(l+s_z)]^2=p_z^2+(2\lambda+1-2s_z)qB+m^2,
\end{equation}
for both the bosons and fermions, where $l$ and $s_z$ are the quantum number of the orbit and spin angular momentum respectively. For unbounded system, $\lambda$ is a non-negative integer and for finite system $\lambda$ depends on the boundary condition. Let us first focus on the unbounded case. The right-hand side of Eq.~(\ref{eq:epB}) is the well-known Landau-level quantization. Rotation enters the dispersion relation by a shift of the energy in the left-hand side, which is expected from the discussion in previous sections that rotation has similar effect with the chemical potential.

At $\Omega=0$, each Landau level is degenerate with the degeneracy factor
\begin{equation}\label{eq:deg}
  N=\Big\lfloor\frac{qBS}{2\pi}\Big\rfloor,
\end{equation}
where $S$ is the area of the $xy$-plane of the system. Thus for $\lambda$th Landau level, $l$ takes integer values in
\begin{equation}
  -\lambda\leqslant l\leqslant N-\lambda,
\end{equation}
and labels the degenerate angular modes of the Landau level $\lambda$. At finite $\Omega$, each Landau level is splitted to $N$ non-degenerate levels separated by $\Omega$. Precisely speaking, $l$ should run up to $N-\lambda-1$, but we consider sufficiently strong magnetic field or large $S$ so that $N\gg 1$, thus we can approximate the upper bound to be $N-\lambda$.

Although here we don't impose any boundary condition on the wave function for the unbounded case, we still have to limit the system size by $R\Omega\leqslant 1$ to preserve the casuality. On the other hand, in order to discuss the Landau quantization in the cylindrical system, the radius $R$ should be larger than the magnetic length $1/\sqrt{qB}$. Therefore, our treatment here is legitimate if $R$ is large enough to ignore the boundary effect on the Landau quantization, but not too large to maintain the causality. That is, the following condition should be imposed:
\begin{equation}
  1/\sqrt{qB}\ll R \leqslant 1/\Omega.
\end{equation}

For simplicity we first assume that the condensate is spatially homogeneous. Following the standard procedure, one can get the thermodynamic potential at zero temperature and chemical potential (for $m_0=0$) \cite{Chen:2015hfc}
\begin{equation}
   V_{\rm eff}=\frac{\sigma^2}{2G}-\frac{qB}{2\pi}\sum_{\lambda=0}^\infty \alpha_\lambda\int^\infty_{-\infty}\frac{dp_z}{2\pi}\sqrt{p_z^2+m_\lambda^2}+V_\Omega,
 \end{equation}
where the rotational contribution is
\begin{equation}\label{eq:Vo}
   V_\Omega=-\frac{1}{S}\sum_{\lambda=0}^\infty \alpha_\lambda\sum_{l=-\lambda}^{N-\lambda}\theta(\Omega|j|-m_\lambda)\int^{k_{\lambda j}}_{-k_{\lambda j}}\frac{dp_z}{2\pi}\left[\Omega|j|-\sqrt{p_z^2+m_\lambda^2}\right]
\end{equation}
with $\alpha_\lambda=2-\delta_{\lambda 0}$, $m_\lambda^2=2\lambda qB+\sigma^2$ and $k_{\lambda j}=\sqrt{(\Omega j)^2-m_\lambda^2}$. The gap equation reads
\begin{equation}\label{eq:gapbo}
\frac{\sigma}{G}=\sigma\frac{qB}{2\pi}\sum_{\lambda=0}^\infty \alpha_\lambda\left[\int^\infty_{-\infty}\frac{dp_z}{2\pi}-\frac{1}{N}\sum_{l=-\lambda}^{N-\lambda}\theta(\Omega|j|-m_\lambda)\int^{k_{\lambda j}}_{-k_{\lambda j}}\frac{dp_z}{2\pi}\right]\frac{1}{\sqrt{p_z^2+m_\lambda^2}}.
\end{equation}
Since the integrand on the right-hand side (RHS) is positive, we observe that the presence of a nonzero $\Omega$ always gives a negative contribution to the RHS and this requires a smaller $\sigma$ (comparing to the $\Omega=0$ case) to balance the left-hand side (LHS). This is consistent with the results in previous sections that the rotation tends to suppress the chiral condensate. Another interesting observation is that the $\Omega$-related terms in gap equation (\ref{eq:gapbo}) is very similar to the gap equation at finite chemical potential, supporting analogy between the rotational and density effects as we discussed before.

It is worthwhile to look at pure magnetic-field effect on the chiral condensate. For simplicity, we consider the strong magnetic-field limit so that lowest-Landau-level (LLL) approximation can apply. Under LLL approximation, the gap equation becomes
\begin{equation}\label{eq:strong}
\frac{\sigma}{G}=\sigma\frac{qB}{2\pi}
\int^\infty_{-\infty}\frac{dp_z}{2\pi}\frac{1}{\sqrt{p_z^2+\sigma^2}}\approx \sigma\frac{qB}{2\pi^2}\ln\left(\frac{2\Lambda}{\sigma}\right) + O\left(\frac{\sigma}{\Lambda}\right)^0,
\end{equation}
where $\Lambda$ is a ultraviolet cutoff for $p_z$. The non-trivial solution to the above equation is $\sigma\approx 2\Lambda\exp[-G_c\Lambda^2/(GqB)]$ with $G_c=2\pi^2/\Lambda^2$ the critical coupling for the onset of chiral condensate at $B=0$. This shows that in the presence of strong magnetic field, no matter how small the coupling $G$ is there is always a nonzero chiral condensate. This phenomenon is called the magnetic catalysis of chiral condensate~\cite{Gusynin:1994re,Gusynin:1995nb}.
\begin{figure}[t]
\begin{minipage}[t]{0.47\linewidth}
\includegraphics[scale=0.13]{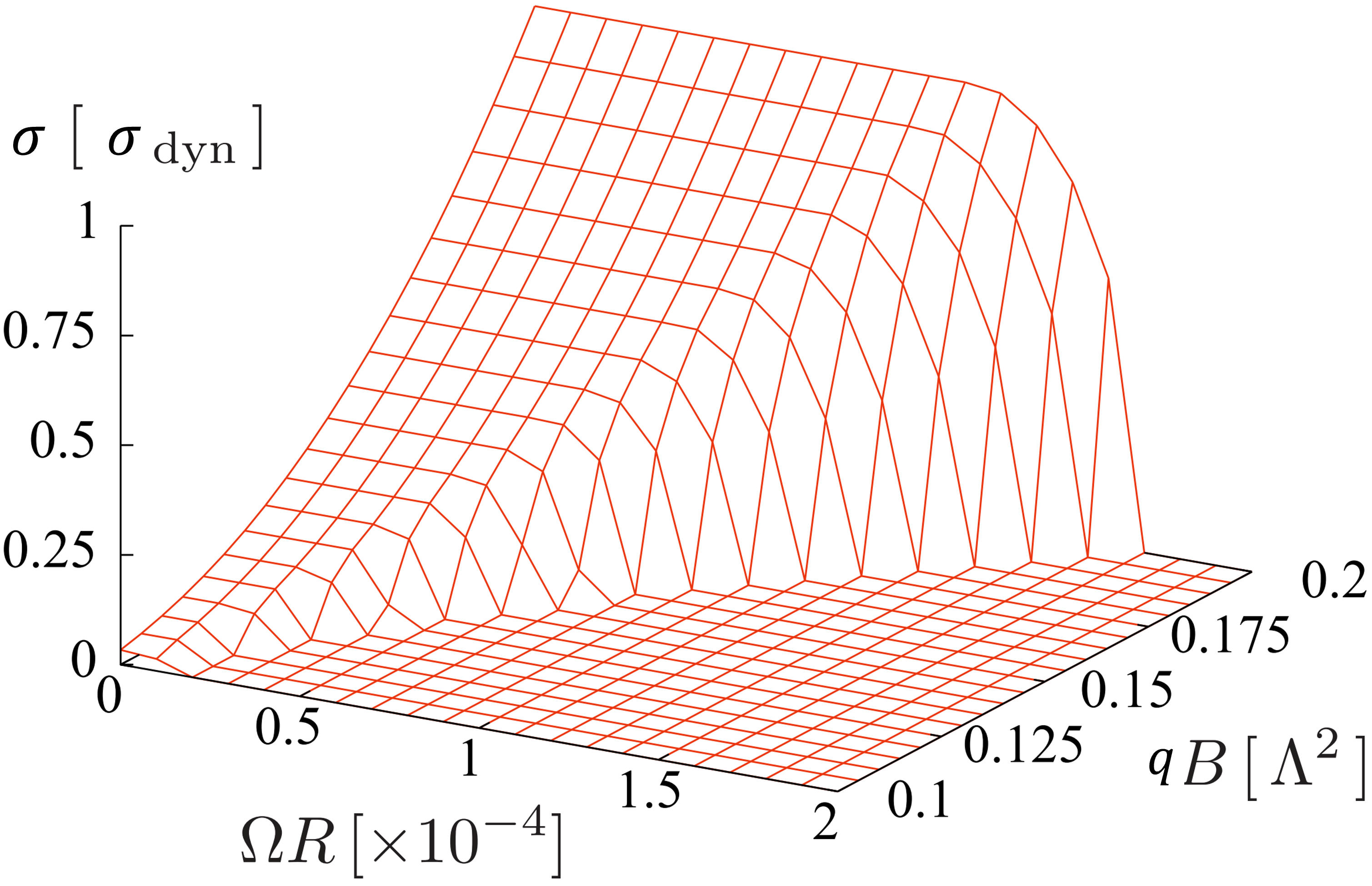}{}
\caption{Chiral condensate as a function of $\Omega$ and
$qB$ at weak coupling, where $\sigma_{\rm dyn}$ is the condensate at $qB=0.2\Lambda^2$ and $\Omega=0$. (Taken from \cite{Chen:2015hfc}.)}
\label{fig:chen2}
\end{minipage}%
\hfill
\begin{minipage}[t]{0.47\linewidth}
\includegraphics[scale=0.13]{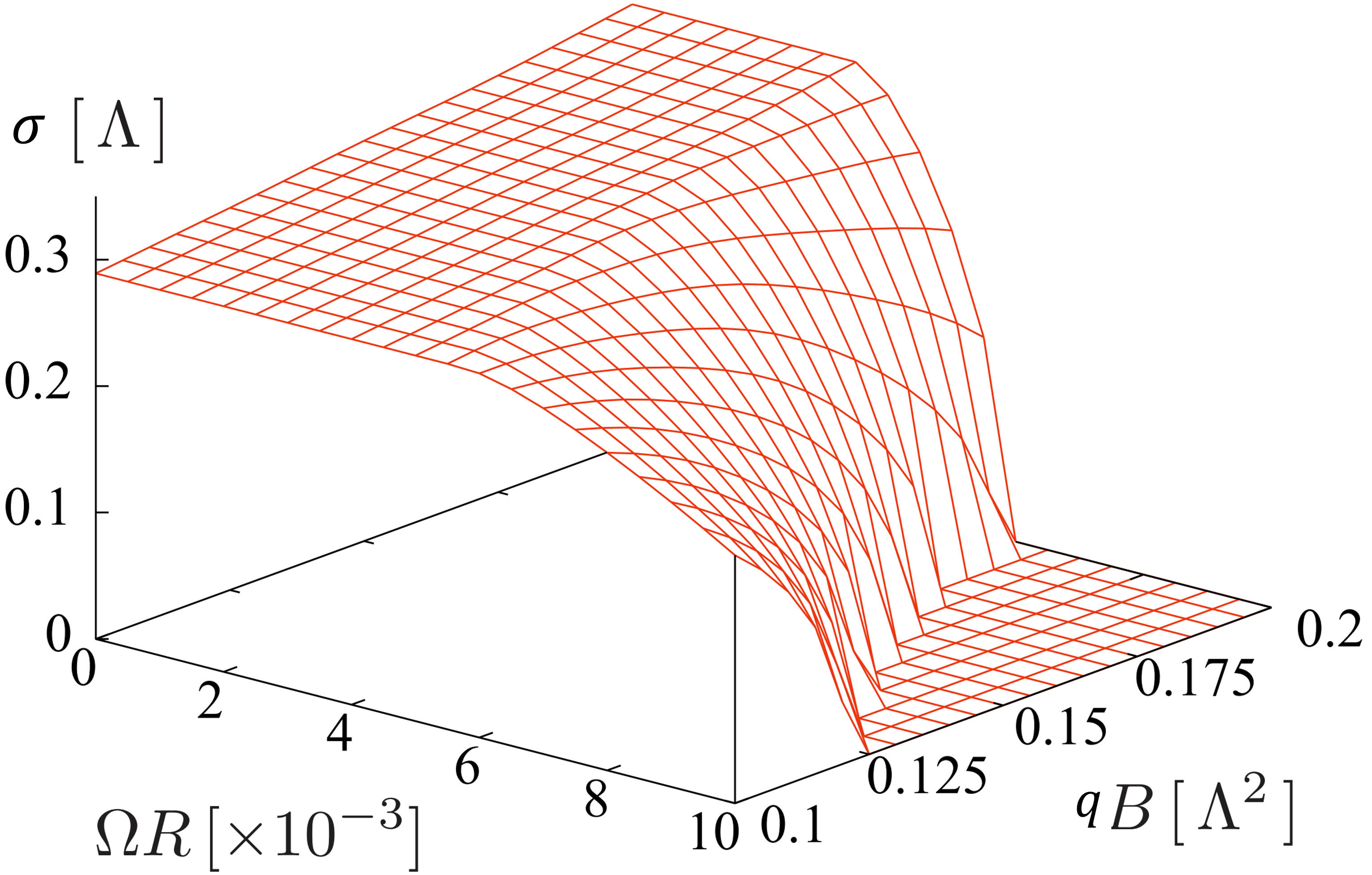}{}
\caption{Chiral condensate as a function of $\Omega$ and
$qB$ at strong coupling. For large $\Omega$, chiral symmetry is restored by
increasing $qB$, which manifests the rotational
magnetic inhibition. (Taken from \cite{Chen:2015hfc}.)}
\label{fig:chen1}
\end{minipage}
\end{figure}

Let us discuss the following two cases separately: (A) $G<G_c$ and (B) $G>G_c$. In the weak coupling case ($G<G_c$), the 3-dimensional plot for the chiral condensate $\sigma$ as a function of $\Omega$ and $qB$ is shown in Fig.~\ref{fig:chen2}~\cite{Chen:2015hfc} where $\sigma_{\rm dyn}$ is the value of the condensate at $qB=0.2\Lambda^2$ and $\Omega=0$. From Fig.\ref{fig:chen2}, we again see that rotation can restore the chiral symmetry. We note that the numerical results in \cite{Chen:2015hfc} are obtained using a smoothed cutoff function in regularizing the $p_z$ integral~\cite{Gorbar:2011ya}: $f(p_z,\Lambda)=\sinh(\Lambda/\delta\Lambda)/[\cosh(\sqrt{p_z^2+2\lambda qB}/\delta\Lambda)+\cosh(\Lambda/\delta\Lambda)]$ with  $\delta\Lambda=0.05\Lambda$.
In the strong coupling case ($G>G_c$), the 3-dimensional plot of $\sigma$ on $\Omega$ and $qB$ is shown in Fig.~\ref{fig:chen1}. For small angular velocity, the chiral condensate is almost independent of $\Omega$ and $qB$. With increasing $\Omega$ the chiral condensate is eventually suppressed by larger magnetic field, i.e. a counterpart of the finite-density inverse magnetic catalysis~\cite{Preis:2012fh} is manifested. This phenomenon is named ``rotational magnetic inhibition"~\cite{Chen:2015hfc}. Similar observation was made also in Ref.~\cite{Liu:2017zhl}.

The difference between the weak and strong coupling cases can be explained by the contribution from the higher Landau levels. In the weak coupling case only a small number of the Landau levels contribute to the gap equation, while many more Landau levels get involved as the coupling constant becomes larger. This is essential for the realization of the rotational magnetic inhibition as well as of the inverse magnetic catalysis at finite density.

One can go further by taking into account the boundary condition. In Ref.~\cite{Chen:2017xrj}, the authors adopt the no-flux boundary condition (\ref{eq:noflux}) and discussed the pure boundary effect with magnetic field but without rotation. With the no-flux boundary condition, $\lambda$ cannot be integer any more, and depends on $l$, which we will denote as $\lambda_{l,k}$. As discussed in Sec.~\ref{sec:bc}, the no-flux boundary condition can not uniquely fix the solution, so we adopt a subsidiary condition~\cite{Chen:2017xrj},
\begin{equation}\label{eq:suff}
  \lambda_{l,k}=
  \Bigg\{
\begin{aligned}
&\xi_{l,k} \quad \text{for} \quad l=0,1,\dots \\
&\xi_{-l-1,k}-l \quad \text{for} \quad l=-1,-2,\dots  \\
\end{aligned}
\end{equation}
where $\xi_{l,k}$ denotes the $k$th zero of the confluent hypergeometric function $ _1F_1(-\xi,l+1,\alpha)$, and $\alpha$ is the dimensionless parameter defined by
\begin{equation}
  \alpha\equiv\frac{1}{2}qBR^2.
\end{equation}
Note that without magnetic field, the system has charge-conjugation ($C$) and CP symmetries, and it is natural to adopt such subsidiary condition which preserves $C$ and CP symmetries. While in the present case $B\neq 0$, $C$ and $CP$ symmetries are explicitly broken by the magnetic filed. Thus we choose Eq.~(\ref{eq:suff}) for the sake of convenience to connect to the $B=0$ limit smoothly.
\begin{figure}[t]
\begin{minipage}[t]{0.47\linewidth}
\includegraphics[scale=0.17]{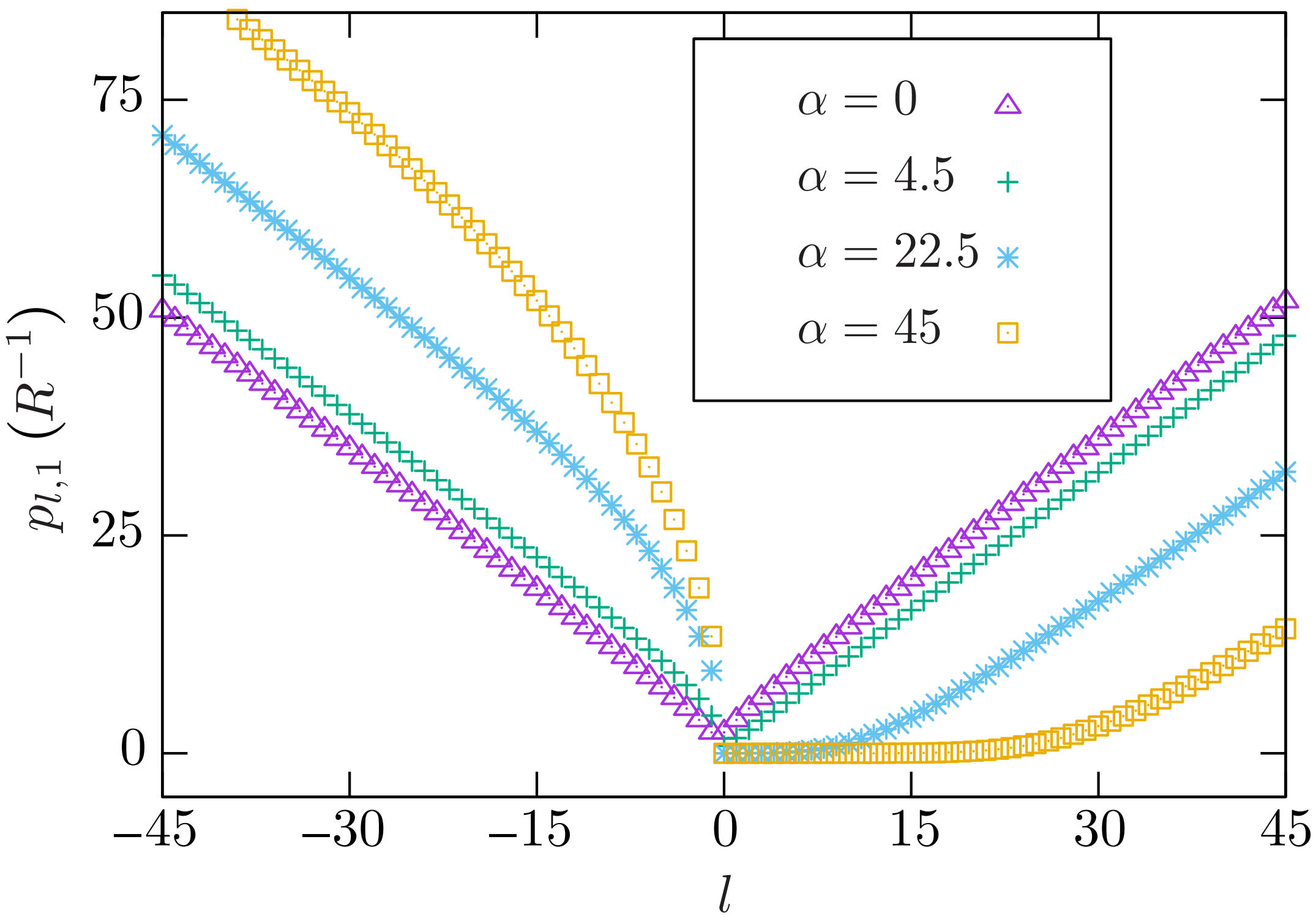}{}
\caption{Lowest transverse momentum $p_{l,1}=\sqrt{2qB\lambda_{l,1}}$ as a function of the angular momentum $l$ for various $\alpha$'s. (Taken from \cite{Chen:2017xrj}.)}
\label{fig:chen5}
\end{minipage}%
\hfill
\begin{minipage}[t]{0.47\linewidth}
\includegraphics[scale=0.137]{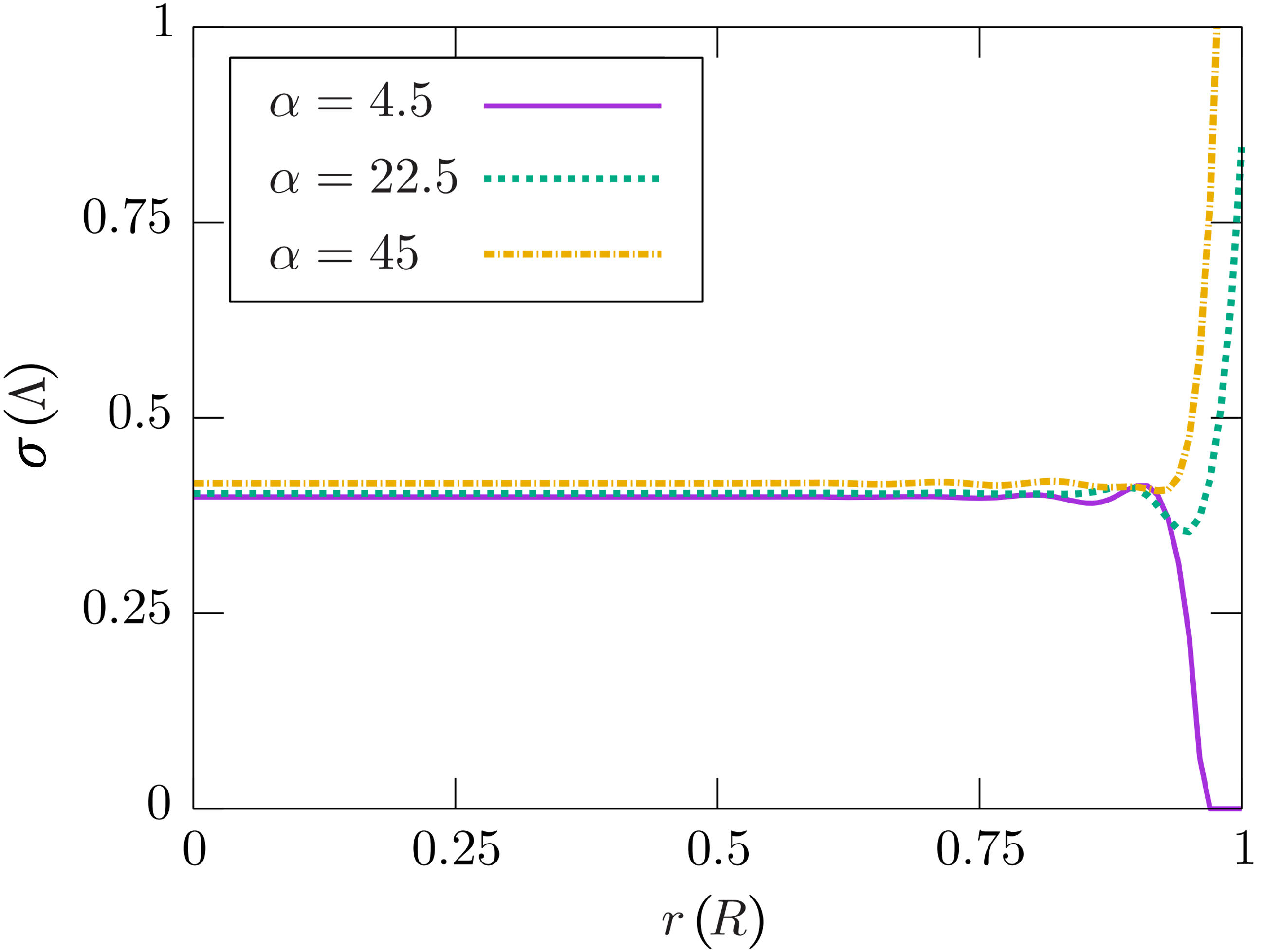}{}
\caption{Inhomogeneous chiral condensate as a function of the radial coordinate $r$ for the choice of $R = 30\Lambda^{-1}$. (Taken from \cite{Chen:2017xrj}.)}
\label{fig:chen6}
\end{minipage}
\end{figure}

In Fig.~\ref{fig:chen5} \cite{Chen:2017xrj} we show the lowest transverse momentum $p_{l,1}=\sqrt{2qB\lambda_{l,1}}$ (this can be viewed as the finite-size modification to the LLL energy) as a function of the angular momentum $l$ for various $\alpha$. In the $B = 0$ case (purple triangular points), positive $l$ modes and negative $(-l-1)$ modes have a degenerated $p_{l,1}$ due to the $CP$ invariance which implies $j \leftrightarrow -j$ symmetry. At finite magnetic field, however, the momenta for the $l > 0$ branch are more suppressed than the $l < 0$ branch because the magnetic field violates $CP$ symmetry and thus a particular direction of the angular momentum is favored. As $\alpha$ increases (i.e. either $B$ or $R$ increases), more $l>0$ modes will be suppressed and finally, we recover the usual Landau zero modes ($\lambda_{0,1}=0$ for $l\geqslant 0$) at $\alpha \to \infty$.

The wave functions with larger $l$ peak at larger $r$ due to the centrifugal force. In the presence of the magnetic field, the wave functions of spin-up and spin-down modes behave differently: with increasing magnetic field the spin-down modes are repelled outward further than the spin-up modes and are eventually accumulated at the boundary. (Note that in the unbounded system, the strong magnetic field would repel the spin-down-mode wave functions to infinity and leave the whole system spin polarized; this is how the LLL dominates at strong magnetic field.) In such a way, the low-energy phenomena closer to the boundary is more prominently affected by the strong magnetic field. One example is the chiral condensate which is a condensate of spin-aligned (but total angular momentum zero) quark-antiquark pairs. In Fig.~\ref{fig:chen6} the chiral condensate solved from the gap equation (Eq.(28) in Ref. \cite{Chen:2017xrj}) in the local density approximation is shown. At small magnetic field, $\alpha = 4.5$, we observe a situation similar to the zero-magnetic field case in comparison with Fig.~\ref{fig:ebihara1}. In contrast to $\alpha= 4.5$, the chiral condensate behavior for stronger magnetic fields ($\alpha= 22.5$ and $45$ in Fig.~\ref{fig:chen6}) is
qualitatively different. Away from the boundary, there is almost no difference. But near the boundary, the chiral condensate is pushed up and its value increases with $\alpha$. This abnormally enhancement is named ``surface magnetic catalysis" in Ref.~\cite{Chen:2017xrj}.

\begin{figure}[t]
\begin{center}
\includegraphics[scale=0.6]{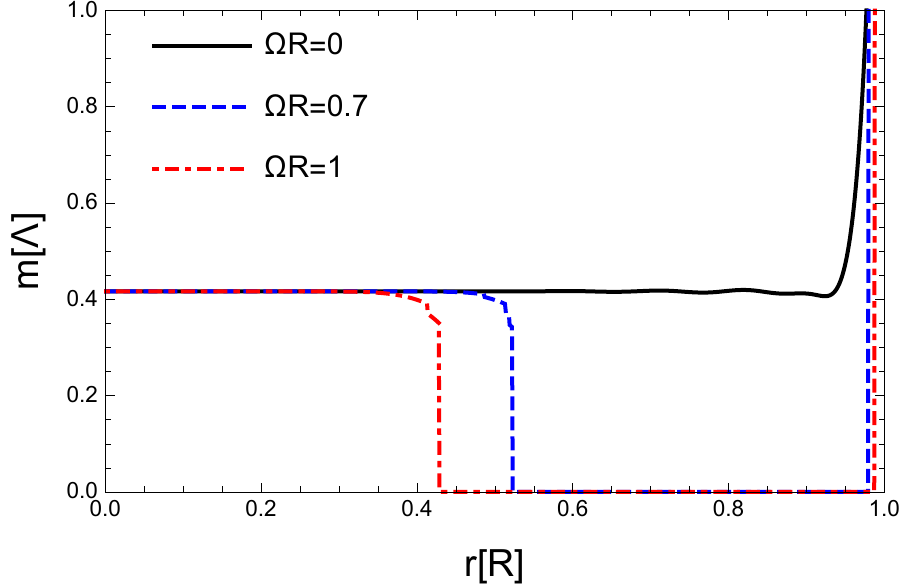}{}
\caption{Inhomogeneous chiral condensate as a function of the radial coordinate
$r$ for different angular velocity at $\alpha=45$. }
\label{fig:chen7}
\end{center}
\end{figure}

Then let us we take rotation into account, the result is shown in Fig.~\ref{fig:chen7}. We can see that the rotational inhibition first occurs near the boundary. The inhibition effect gets closer to the center with the increasing of the angular velocity. Due to the restriction of casuality, the angular velocity can not be very large, thus the condensate at the center of the cylinder will not be inhibited. We can also observe that there is a competition between rotational magnetic inhibition and surface magnetic catalysis at the region very close to the boundary~\cite{Chen:2019tcp}. (Because the chiral condensate in the immediate vicinity of the boundary is strongly inhomogeneous, the local density approximation could break down. But the surface magnetic catalysis is expected to be qualitatively unchanged as it comes from a number of accumulating modes at $r\approx R$; see Refs. \cite{Chen:2017xrj,Chen:2019tcp} for more discussions.)

\section{Inhomogeneity of Chiral Condensate: A BdG treatment}
During the above discussion, the local density approximation is adopted to deal with inhomogeneous condensate, which assumes that the condensate varies slowly with position, i.e. $\partial\sigma\ll\sigma^2$. As we can see from Fig.\ref{fig:ebihara1}, Fig.\ref{fig:chen6}, and Fig. \ref{fig:chen7}, the local density approximation breaks down in the vicinity of the boundary. A more realistic way to handle the inhomogeneous condensate is to solve the Bogoliubov de Gennes (BdG) equation, which has been utilized in \cite{Wang:2018zrn} and \cite{Wang:2019nhd} in the discussion of vortex in chiral condensate and also of the non-uniform rotation. 

The BdG method amounts to solve the mean-field gap equations and the eigenvalue problem of the Hamiltonian self-consistently. Let us recall the Dirac Hamiltonian $\hat{H}[\sigma(\bm x), \pi(\bm x)]$ (see Eq. \ref{eq:Hamiltonian};  the authors of \cite{Wang:2018zrn} and \cite{Wang:2019nhd} considered the case with one flavor and no background gauge field) and the thermodynamic potential $V_{\rm eff}[\sigma(\bm x), \pi(\bm x)]$ (see Eq. \ref{therml}). The BdG equation is just the eigenequation of $\hat{H}$ with appropriate boundary conditions: $\hat{H}\Psi_{\{\xi\}}(\bm{x})=\varepsilon_{\{\xi\}}\Psi_{\{\xi\}}(\bm{x})$ where $\Psi_{\{\xi\}}(\bm{x})$ and $\varepsilon_{\{\xi\}}$ depend on the profile of $\sigma(\bm x)$ and $\pi(\bm x)$. Supplemented by the gap equations $\delta V_{\rm eff}/\delta\sigma(\bm x)=\delta V_{\rm eff}/\delta\pi(\bm x)=0$, the BdG equation determines the condensates and the corresponding wave functions.

\begin{figure}[t]
\begin{minipage}[t]{0.47\linewidth}
\includegraphics[scale=0.13]{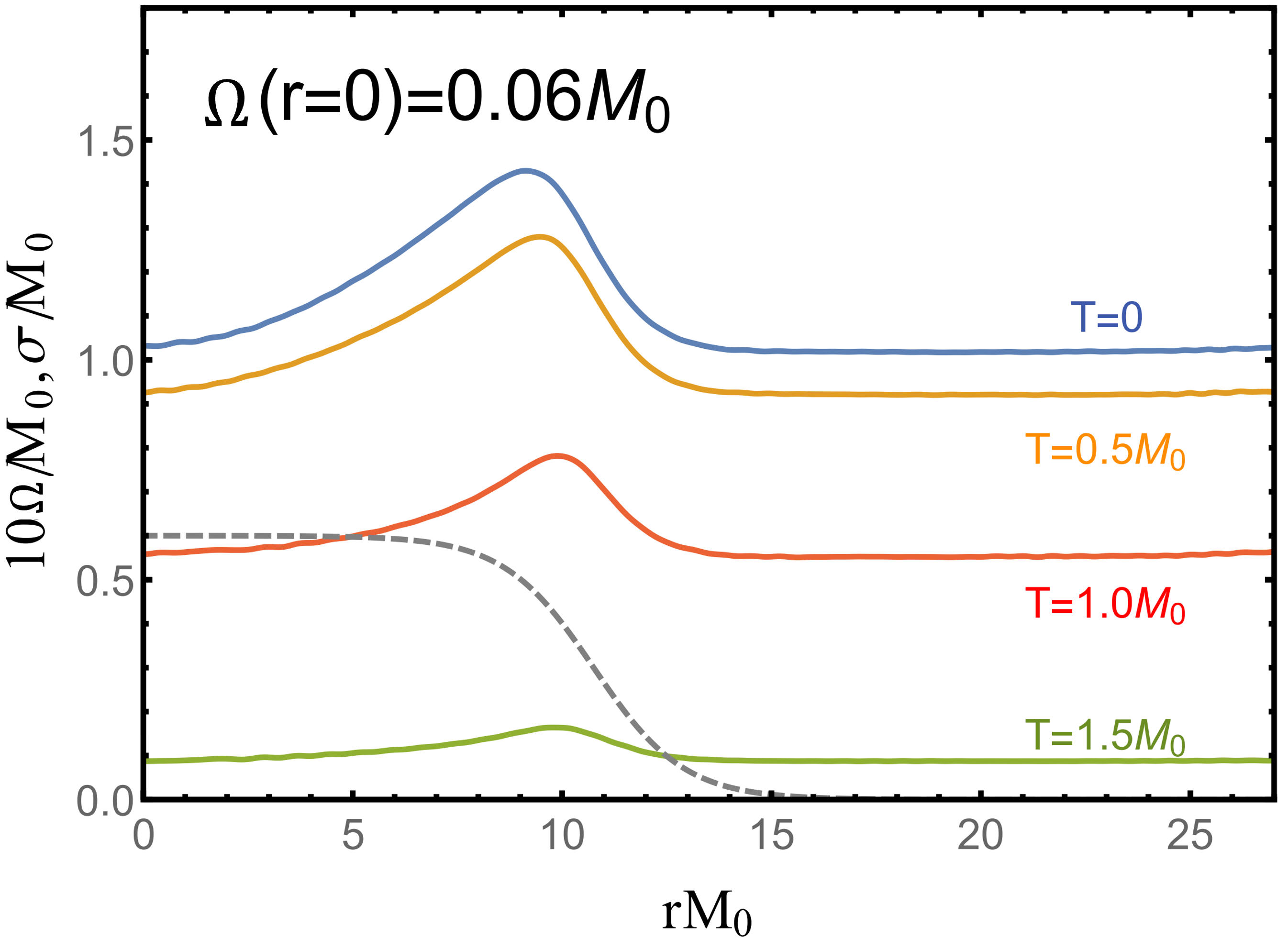}{}
\caption{Chiral condensate profile $\sigma(r)$ (solid lines) at different
temperatures under rotation $\Omega = 0.06M_0[\exp(r M_0 - 10) +
1]^{-1}$ (dashed line). Here $M_0$ is the chiral condensate in infinite system at $T=\Omega=0$. (Taken from \cite{Wang:2018zrn}.)}
\label{fig:wang1}
\end{minipage}%
\hfill
\begin{minipage}[t]{0.47\linewidth}
\includegraphics[scale=0.134]{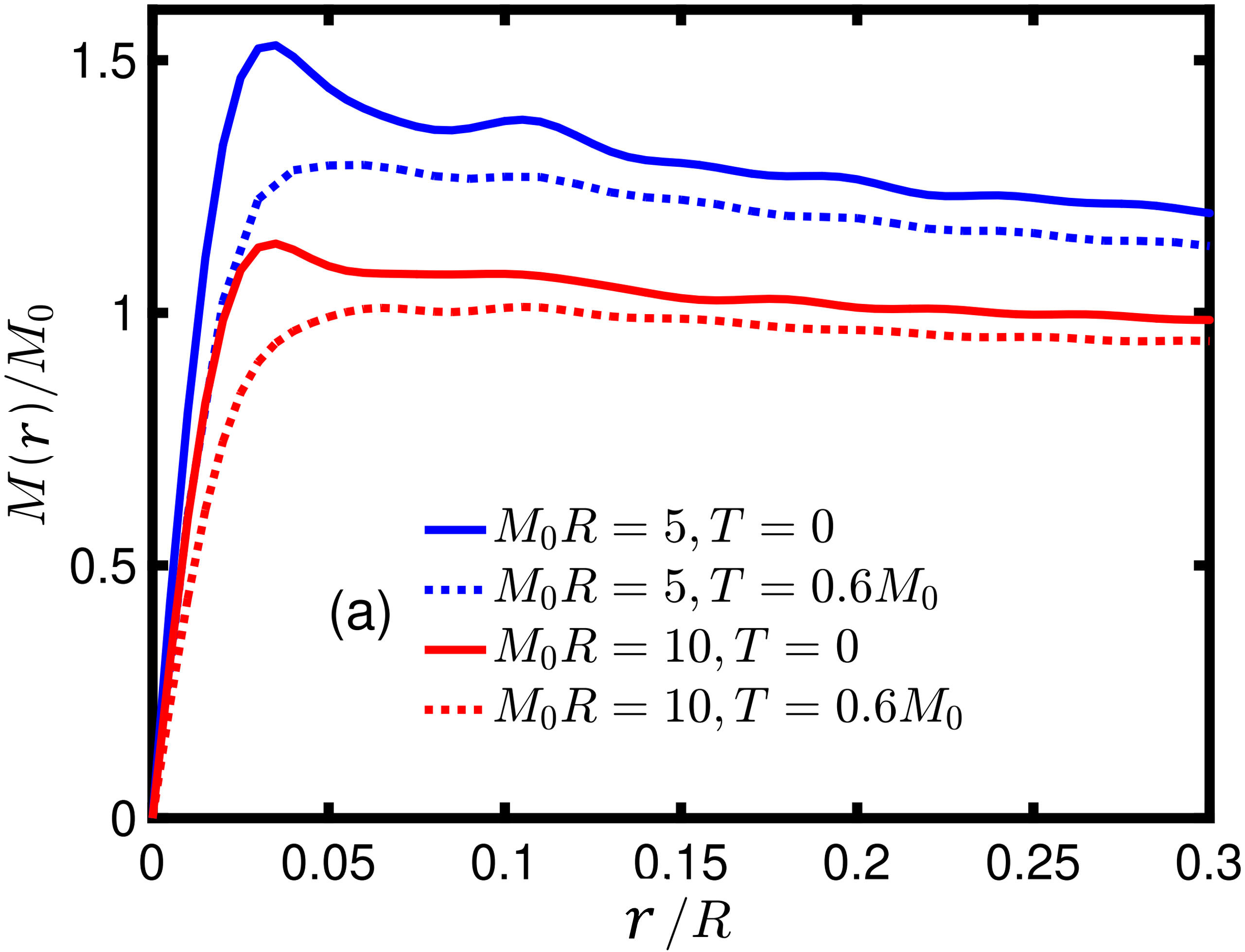}{}
\caption{Chiral condensate with a vortex excitation of $\kappa=1$ for different values of $R$ and $T$ at $\Omega=0$. (Taken from \cite{Wang:2019nhd}.)}
\label{fig:wang2}
\end{minipage}
\end{figure}

In Ref.~\cite{Wang:2018zrn}, the authors studied the 2+1D NJL model rather than 3+1D, since the NJL model is renormalizable in 2+1D. The authors studied the chiral condensate under non-uniform rotation. Choosing a Woods-Saxon shaped rotation profile $\Omega(r)=\Omega_0/[\exp(r-r_0)+1]$, the BdG method gives the transverse profile of the chiral condensate as shown in Fig.~\ref{fig:wang1}. Since the angular velocity $\Omega_0$ shown in Fig.~\ref{fig:wang1} is small, the rotational suppression effect is not evident. Instead, the chiral condensate $\sigma$ increases slowly in the range of the rotation plateau and falls back to $\sigma(\Omega=0)$ rapidly beyond the rotational region. A bump structure appears around $r_0$ which reflects the large gradient of $\Omega_0$ around $r_0$. Physically, the bump structure may be due to the rotational energy shift $\Delta E(r)=-\Omega(r) J_z$ which results in an angular-momentum dependent radial force $F(x)=-\partial_r \Delta E(r)=\partial_r\Omega(r)J_z$ playing an analogous role as the centrifugal force.

It is natural to consider the possible vortex structure induced by rotation by employing the BdG method. In Ref.~\cite{Wang:2019nhd}, the authors studied the one flavor NJL model in 2+1D rotating spacetime with a constant angular velocity $\Omega$. Due to the $U(1)$ chiral symmetry of the system, one can define a complex order parameter
\begin{equation}
  \Delta(x)=\sigma(x)+i\pi(x)=M(x)e^{i\phi(x)},
\end{equation}
where $M$ and $\phi$ are set to be real. The phase $\phi(x)$ is chosen to be
\begin{equation}
  \phi=i\kappa \theta, \quad \kappa \in \mathbb Z,
\end{equation}
with $\theta$ the azimuthal angle. Our previous discussions only cover the case with $\kappa=0$. The case with $\kappa\neq 0$ corresponds to the quantized vortex state. Solving the BdG equation with the vortex ansatz can give us the vortex structure. If the condensate with such a vortex provides lower thermodynamic potential than the one without the vortex (i.e. for the case of $\kappa=0$), it means that the vortex structure is thermodynamically favored.

The vortex core structure with $\kappa=1$ is shown in Fig.~\ref{fig:wang2}~\cite{Wang:2019nhd}. The structure is expected to be qualitatively similar for finite $\Omega$. For $\Omega=0$ case, the vortex corresponds to an excited state and is not thermodynamically stable. But since the vortex carries a finite angular momentum, it is more favorable when the system is rotating. Indeed, one can examine the difference in thermodynamic potential, $\delta V_{\rm eff} = V_{\kappa=0} - V_{\kappa=1}$ to determine the favored thermodynamical state, and when $\Omega$ exceeds a critical value $\Omega_c$, $\delta V_{\rm eff}$ becomes positive meaning that the vortex state becomes the more stable. For large system size, the critical angular velocity exceeds the causality bound (i.e. $\Omega_cR>1$), thus it is impossible for a vortex state to stably exist~\cite{Wang:2019nhd}.

\section{Mesonic Superfluidity}

\begin{figure}[t]
\begin{center}
\includegraphics[scale=0.14]{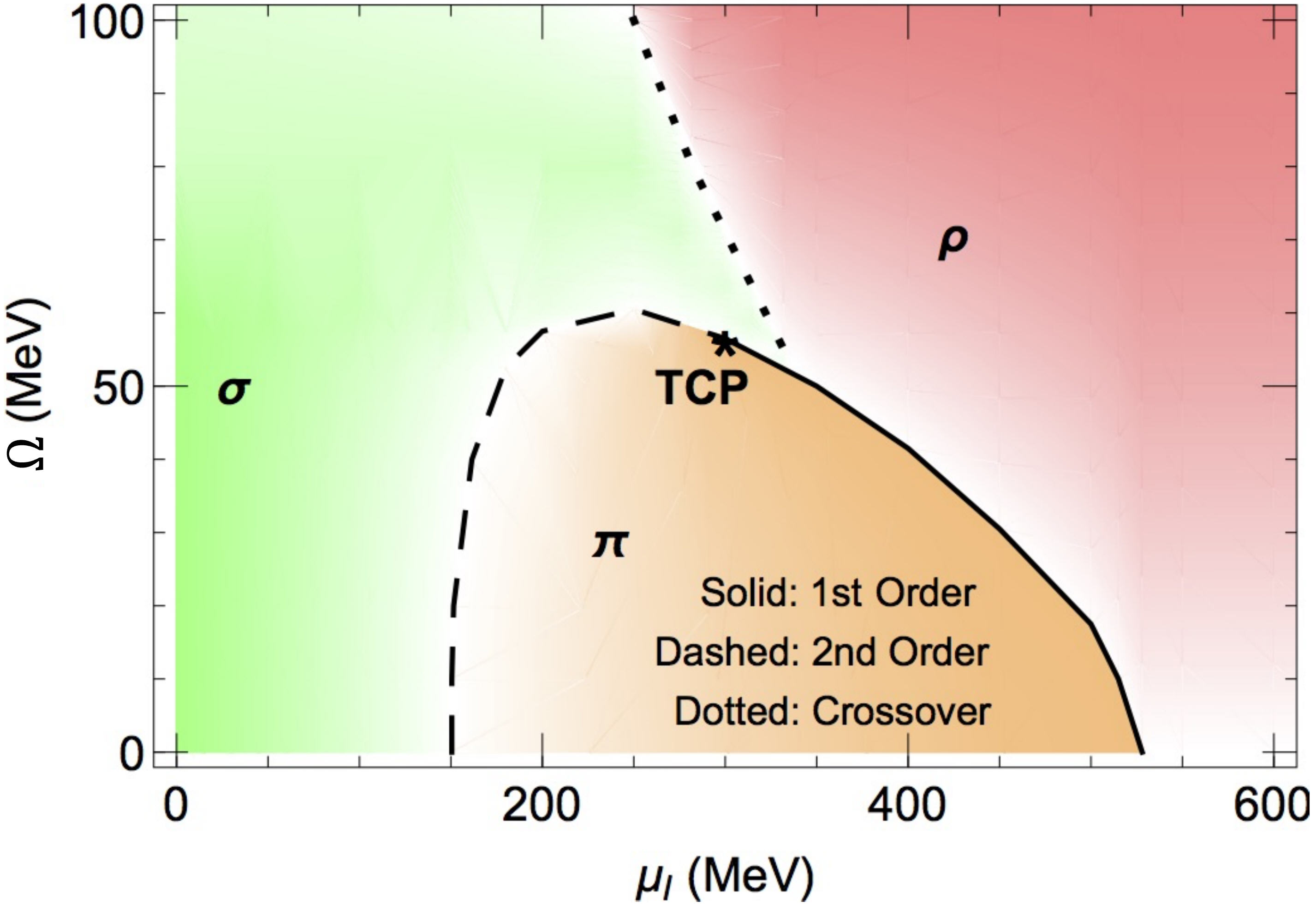}{}
\caption{Phase diagram on the $\Omega-\mu_I$ plane for mesonic superfluidity in isospin matter under rotation. (Taken from \cite{Zhang:2018ome}.)}
\label{fig:hou}
\end{center}
\end{figure}

We have focused on the chiral condensate in the previous sections. In this section we discuss the condensation of other types of pairings, particularly the pions. Such condensate usually triggers a superfluidity at finite density, so in the following we will not distinguish the term ``condensate" and the term ``superfluid". As pions are $J=0$ mesons, it is clear that the pion condensation would not be favored by uniform rotation. On the other hand, as $J=1$ meson, the rho condensation would be favored by uniform rotation~\cite{Zhang:2018ome,Cao:2020pmm}. In Ref.~\cite{Zhang:2018ome}, the authors took into account the isospin chemical potential in NJL model and considered the rotational effect on the pion and rho condensates. The Lagrangian is the sum of two-flavor NJL Lagrangian~(\ref{eq:NJL}) (with $A_\mu=0$), ${\mathcal L}_I=(\mu_I/2)\bar{\psi}\gamma^0\tau_3\psi$, and ${\mathcal L}_\rho=-(G_\rho/2)(\bar\psi\gamma_\mu\bm{\tau}\psi)^2$. Considering the unbounded case and the local density approximation to the condensates $\sigma=-G\langle{\bar\psi}\psi\rangle$, $\pi=-G\langle{\bar\psi}i\gamma_5\tau_3\psi\rangle$, and $\rho=-G_\rho\langle{\bar\psi}i\gamma_0\tau_3\psi\rangle$, the thermodynamic potential can be derived in a similar manner as described in Sec.~\ref{sec:unbound}. The result is $V_{\rm eff}=\int d^3{\bm r}{\mathcal V}_{\rm eff}(r)$ with
\begin{equation}
\begin{split}\label{eq:tp}
{\mathcal V}_{\rm eff}(r)&=\frac{\sigma^2+\pi^2}{2G}-\frac{\rho^2}{2G_\rho}-\frac{N_fN_c}{16\pi^2}\sum_{a=\pm}\sum_l\int dp_t^2\int dp_z\\
  &\;\;\;\;\;\times[J_l(p_tr)^2+J_{l+1}(p_tr)^2] T\ln(1+e^{\beta\varepsilon^a_{l}})(1+e^{-\beta\varepsilon^a_{l}}),
\end{split}
\end{equation}
where $\varepsilon^\pm_{l}=\sqrt{\pi^2+(\sqrt{m^2+p_t^2+p_z^2}\pm\tilde{\mu}_I/2)^2}-\Omega(l+1/2)$, $m=m_0+\sigma$, $\tilde{\mu}_I=\mu_I+G_\rho\rho$, and $N_f=2, N_c=3$. The thermodynamically equilibrium state is specified by the minimum of $V_{\rm eff}$. As $\Omega$ and $\mu_I$ vary, the true equilibrium state would vary as well leading to phase transitions. The phase diagram so obtained is plotted in Fig.~\ref{fig:hou} (see Ref. \cite{Zhang:2018ome} for the parameter setup) which is characterized by three distinctive regions: a vacuum-like sigma-dominated phase in the low isospin chemical potential and slow rotation region, a pion-superfluid phase in the mid-to-high isospin density with moderate rotation, and a rho-superfluid phase in the high isospin and rapid rotation region. A second-order transition line separates the sigma-dominant and the pion-dominant regions while a first-order transition line separates the pion-dominant and the rho-dominant regions, with a tri-critical point (TCP) connecting them. Note that without electromagnetic interactions, there is no distinction among the isospin-1 triplet states $\pi^{0,\pm}$ or among the $\rho^{0,\pm}$ and one is free to choose the condensation to be in any direction of the isospin space.

In the above discussion we considered rotating isospin matter. Adding baryon charges into such isospin matter would induce intriguing phenomenon through chiral anomaly. The underlying mechanism can be understood through the delicate chiral vortical effect (CVE) for axial current $j^{a\mu}_5=\langle\bar{\psi}\gamma^\mu\gamma_5\tau^a\rangle$. Consider the case of $a=3$ for two-flavor quarks at $T=0$. The CVE current is given by
\begin{equation}\label{cve}
   \bm{j}^3_5=\frac{\mu_B\mu_I}{\pi^2}\bm{\Omega}.
\end{equation}
This current is protected by chiral anomaly and is thus exact regardless of the energy scale. Therefore we can write down a low-energy effective Hamiltonian for the CVE current (\ref{cve}) in terms of pion field~\cite{Huang:2017pqe}:
\begin{equation}\label{effectivecve}
   {\mathcal H}_{\rm CVE}=-\frac{\mu_B\mu_I}{2\pi^2 f_\pi}\bm{\nabla}\pi_0\cdot\bm{\Omega},
\end{equation}
where $f_\pi$ is pion decay constant. Adding the kinetic terms for $\pi_0$ field, we get the total effective Hamiltonian as $H=\int d^3\bm x{\mathcal H}$ with ${\mathcal H}$ being (suppose $\bm\Omega$ to be along $z$ axis)
\begin{equation}\label{effectiveL}
   {\mathcal H}=\frac{1}{2}\left[(\partial_r\pi_0)^2+\frac{1-(\Omega r)^2}{r^2}(\partial_\theta\pi_0)^2+(\partial_z\pi_0)^2\right]+m_\pi^2f_\pi^2
   \left(1-\cos\frac{\pi_0}{f_\pi}\right)-\frac{\mu_B\mu_I}{2\pi^2 f_\pi}\Omega\partial_z\pi_0,
\end{equation}
where $m_\pi$ is pion mass. The ground state must minimize $H$ which is specified by the equations of motion for $\pi_0$: $\partial_r\pi_0=\partial_\theta\pi_0=0$ and $\partial_z^2\pi_0=m_\pi^2 f_\pi\sin(\pi_0/f_\pi)$. The solution to the equations of motion is given by~\cite{Huang:2017pqe}
\begin{equation}\label{csl}
\cos\frac{\pi_0(\bar{z})}{2f_\pi}={\rm sn}(\bar{z},k),
\end{equation}
where ${\rm sn}(\bar{z},k)$ is the Jacobi elliptic function of modulus $k\in[0,1]$ and $\bar{z}=m_\pi z/k$. This solution describes an interesting chiral soliton lattice structure of $\pi_0$ condensate along $\bm\Omega$ direction. In each lattice cell, the angular momentum, baryon, and isospin charges are topological quantities (namely, independent of the shape of $\pi_0(z)$). Similar chiral-anomaly induced neutral pion condensate occurs also in parallel electric and magnetic fields~\cite{Cao:2015cka,Wang:2018gmj} and in external magnetic field with high baryon chemical potentials in which the chiral soliton lattice can also appear~\cite{Brauner:2016pko}. Also, similar chiral soliton lattice structure can also be realized in $\eta'$ condensate in rotating baryonic matter~\cite{Nishimura:2020odq}. We note that substituting solution (\ref{csl}) into $H$ one can find that only when $\Omega$ exceeds a critical value $\Omega_c=8\pi m_\pi f_\pi^2/(\mu_B|\mu_I|)$ the chiral soliton lattice corresponds to the lowest energy and is the true ground state of QCD. On the other hand, $\Omega$ must be much smaller that the QCD scale to guarantee the availability of the effective theory (which also guarantees that the rotational suppression of $\pi_0$ condensate is negligible); see more discussions in Ref.~\cite{Huang:2017pqe}.

We have discussed the novel effects of rotation on the neutral pion condensate at finite isospin and baryon chemical potentials. In the following we discuss the effect of a magnetic field on the rotating pionic matter which is related to charged pion superfluid. This was first considered in Refs.~\cite{Liu:2017spl} and \cite{Liu:2017zhl}. By solving the Klein-Gordon equation (\ref{eq:KGeq}), one can get the dispersion relation for charged pion $\pi^\pm$
\begin{equation}
   E=\sqrt{|qB|(2n+1)+p_z^2+m_\pi^2}-{\rm sgn}(q)\Omega l,
 \end{equation}
where $q=e>0$ for positively charged pions and $q=-e$ for negatively charged pions. It is easy to see that the degeneracy of each Landau level is lifted. In particular, the $\pi^+$ in the LLL splits down, and the $\pi^-$ in the LLL splits up. Thus effectively the $\Omega l$ play the role of an isospin chemical potential $\mu_l=\Omega l$ for $\pi^+$ and $-\mu_l=-\Omega l$ for $\pi^-$. Therefore, when $\mu_N=N\Omega$ ($N=\lfloor|qBS|/(2\pi)\rfloor$ is the Landau degeneracy) exceeds the effective mass of $\pi^+$ in the LLL, $m_0=\sqrt{eB+m_\pi^2}$, but it is still below the $\pi^+$ effective mass in the first LL, the LLL $\pi^+$ may Bose-condense. As $\Omega$ increased, higher Landau-level $\pi^+$ may Bose-condense sequentially. Again, we emphasize that we are considering an unbounded system with the constraint $\Omega\leqslant 1/R\ll\sqrt{eB}$ implicitly assumed: the first inequality is due to the causality and the second one is to make the Landau quantization sensible.

Recently, such possible charged pion condensation was re-examined by using the NJL model~\cite{Chen:2019tcp,Cao:2019ctl}. This is important because $\pi^+$ comprises a $u$ quark and a $\bar{d}$ antiquark with their angular momenta (both the spin and the orbital one) antiparallel to each other. Thus both the  magnetic field and rotation would tend to suppress $\pi^+$ condensate. In addition, the rotational magnetic inhibition discussed in Sec.~\ref{sec:OmegaB} would take place which may also influence the charged pion condensate. So we need to re-consider the charged pion condensate from the quark-level dynamics. The Lagrangian is Eq.~(\ref{eq:NJL}). Since the $SU(2)$ isospin symmetry is explicitly broken to $U(1)_{I_3}$ by the background magnetic field, it is very difficult to diagonalize the Hamiltonian Eq. (\ref{eq:Hamiltonian}) with nonzero charged $\pi^\pm=\pi^1\mp \pi^2$ condensates. One can instead consider a Ginzburg-Landau approach for the $\pi^\pm$ fields but with the $\sigma$ field taken into account self-consistently through gap equation. This amounts to examine the stability of the $\sigma$-condensed phase against the onset of charged pion condensate. For this purpose, we expand the thermodynamic potential in $\mathbf\pi$,
\begin{equation}
\label{eq:GammaExpansion}
  \begin{split}
  V_{\rm eff}
  	& =V_{\rm eff}^{(0)}+V_{\rm eff}^{(2)}+\dots\,,\\
  V_{\rm eff}^{(0)}
    &=\frac{1}{\beta V}\int d^4 x_{\rm E} \frac{\sigma^2}{2G}
    -\frac{1}{\beta V}{\rm Tr}\ln(i\slash\!\!\!\nabla+\mu_B\gamma^0-\sigma)\,,\\
  V_{\rm eff}^{(2)}
  	&=\frac{1}{\beta V}\int d^4 x_{\rm E} \frac{\bm{\pi}^2}{2G}
  		-\frac{1}{2\beta V}{\rm Tr}[(i\slash\!\!\!\nabla+\mu_B\gamma^0-\sigma)^{-1}\gamma^5\bm{\pi}\cdot\bm{\tau}]^2\,,
  \end{split}
\end{equation}
where a baryon chemical potential $\mu_B$ is introduced. Note that we set $\pi^0=0$ since we only focus on the charged pion fields.

In Eq. (\ref{eq:GammaExpansion}), $V_{\rm eff}^{(0)}$ has been discussed plentifully in previous sections. What we are interested in is $V_{\rm eff}^{(2)}$ which is quadratic in $\bm\pi$. If we write
 \begin{equation}
  V_{\rm eff}^{(2)}=\int d^4x'_{\rm E}d^4x_{\rm E} \; \pi^+({x'})\;C^{(2)}(x',x)\;\pi^-({x}),
\end{equation}
where $C^{(2)}(x',x)$ is the inverse pion propagator at one-quark-loop level. The charged pion condensation would be favored if $C^{(2)}(x',x)$ is not semi-positive definite (as a matrix). In Refs.~\cite{Chen:2019tcp,Cao:2019ctl}, the sign of $C^{(2)}(x',x)$ is examined by taking the following ansatz to simplify $V_{\rm eff}^{(2)}$:
\begin{equation}\label{asatz}
  \pi^+(x')\pi^-(x)=e^{ie\int^{x}_{x'} A_\mu d z^\mu}\tilde\pi^+\tilde\pi^-,
\end{equation}
where $\tilde\pi^+$ and $\tilde\pi^-$ are gauge-independent condensates which are assumed to be constants and  the exponential factor is the Wilson line which connects $x'$ and $x$. This Wilson line should be so chosen that the Schwinger phase in the quark propagator entering $C^{(2)}(x',x)$ is canceled and hence $V_{\rm eff}^{(2)}=C^{(2)}\tilde\pi^+\tilde\pi^-$ is gauge invariant with the constant $C^{(2)}=\int d^4x'_{\rm E}d^4x_{\rm E} \;C^{(2)}(x',x)e^{ie\int^{x}_{x'} A_\mu d z^\mu}$ playing the role of the Ginzburg-Landau coefficient. The Schwinger phase $\Theta(x,x')$ is the gauge dependent factor in the quark propagator, $S(x,x')=e^{i\Theta(x,x')}S_{\rm inv}(x,x')$ so that $\Theta(x,x)=0$ is satisfied and $S_{\rm inv}$ is gauge invariant. For constant electromagnetic field in rotating frame, it can be shown that the Schwinger phase is given by $\Theta(x,x')= -q\int_x^{x'} A_\mu(z)d z^\mu$ ($q$: the quark charge) along the geodesic between $x$ and $x'$. The numerical study in Ref.~\cite{Chen:2019tcp} shows that under reasonable parameter choice, the Ginzburg-Landau coefficient is positive at $\mu_B=0$ disfavoring the onset of charged pion condensate. But once a large negative $\mu_B$ is supplied, there can indeed be a region in $\Omega$ and $B$ so that the charged pion condensation is favored. On the other hand, if one insists using another integral path away from the geodesic, one can indeed find
charged pion condensate even at $\mu_B=0$~\cite{Chen:2019tcp,Cao:2019ctl}. Finally, we want to emphasize that even one can rule out the possibility of homogeneous charged pion condensates $\tilde\pi^\pm$ as given in the ansatz (\ref{asatz}), it remains still the possibility of an inhomogeneous condensates to be favored. Besides, as the charged pion condensate triggers an electric superconductivity, the Meissner effect would repel the magnetic field from the bulk superconductor and only allow inhomogeneous magnetic vortices. These new possibilities demand future study.

\section{Summary}

In this article we have presented an overview of the recent progress on the understanding of   QCD phase structure under rotation. The rotation plays a role of polarizing the spin (and orbital motion) of underlying microscopic particles and in turn induces a number of novel effects on various condensates of quark-(anti)quark pairs. An enriched phase diagram in the $T$-$\mu_B$-$\Omega$ space is sketched in Fig.~\ref{fig:illu}. At low $T, \mu_B$, and $\Omega$ we have the usual hadronic matter with the spontaneous  chiral symmetry  breaking. At high $T$, either for small or large $\Omega$, the quarks and gluons are expected to be liberated from the hadrons and form a deconfined quark-gluon plasma phase with restored chiral symmetry. At small $T$ and $\Omega$ but asymptotically high $\mu_B$, QCD is in the color superconducting phase with merely spin-0 quark-quark pairings while at moderate $\mu_B$ the ground state of QCD is to a large content unknown. At high $\mu_B$  with increasing $\Omega$, it is plausible to expect  that QCD possibly undergoes a transition from spin-0 to spin-1 color superconductor.
Now with a new dimension of rotation, the QCD phase diagram becomes even more interesting and rich. The rotation is found to considerably influence the chiral condensate. Adding a magnetic field or isospin chemical potential would lead to additional structures through condensations in various mesonic channels. We have discussed recent results on  these effects in great detail by using effective models in rotating frame.
To summarize,  the properties and phase structures of QCD matter under rotation is an interesting emerging direction with many new theoretical questions still to be explored and answered. It is also important to investigate potential implications and observable effects for the QCD system with large vorticity  in heavy-ion collisions as well as the nuclear matter inside   fast rotating compact stars. One would anticipate a lot of exciting progress to be made in the near future.

\section*{Acknowledgements}
We thank Kenji Fukushima, Defu Hou, Kazuya Mameda, Kentaro Nishimura, Yin Jiang, Naoki Yamamoti, Hui Zhang for collaborations. This work is supported in part by the NSFC Grants No.~11535012 and No.~11675041, as well as by the NSF Grant No. PHY-1913729 and by the U.S. Department of Energy, Office of Science, Office of Nuclear Physics, within the framework of the Beam Energy Scan Theory (BEST) Topical Collaboration.

\end{document}